\documentclass[runningheads]{svmult}

\usepackage{makeidx}   
\usepackage{graphicx}  
\usepackage{subeqnar}  
\usepackage{multicol}  
\usepackage{physprbb}  
\usepackage{amssymb}
\usepackage{amsmath}
\makeindex             

\font\tenscr=rsfs10 scaled1100
\font\sevenscr=rsfs7 
\font\fivescr=rsfs5 
\skewchar\tenscr='177 
\skewchar\sevenscr='177
\skewchar\fivescr='177 
\newfam\scrfam 
\textfont\scrfam=\tenscr
\scriptfont\scrfam=\sevenscr 
\scriptscriptfont\scrfam=\fivescr

\def\scri{{\fam\scrfam I}} 

\font\SYM=msbm10 
\newcommand{\Real}{\mbox{\SYM R}}
\newcommand{\Complex}{\mbox{\SYM C}}
\newcommand{\Natural}{\mbox{\SYM N}}
\newcommand{\Mink}{\mbox{\SYM M}}

\def\phihat{\widehat{\phi}}

\def\O{\mathcal{O}}

\begin{document}


%
\title*{Polyhomogeneous expansions close to null\protect\newline and
spatial infinity}
\toctitle{Polyhomogeneous expansion close to null and spatial infinity}

\titlerunning{Polyhomogeneous expansions}
%
\author{Juan Antonio Valiente Kroon \thanks{E-mail address:
{\tt jav@aei-potsdam.mpg.de}}}
\authorrunning{J.A. Valiente Kroon}
%
%
\institute{Max Planck Institute f\"{u}r Gravitationsphysik, \\
           Albert Einstein Institut, \\
           Am M\"{u}hlenberg 1, \\
           14476 Golm bei Potsdam, \\
           Germany}
\maketitle              

\begin{abstract}
  A study of the linearised gravitational field (spin 2 zero-rest-mass
  field) on a Minkowski background close to spatial infinity is done.
  To this purpose, a certain representation of spatial infinity in
  which it is depicted as a cylinder is used. A first analysis shows
  that the solutions generically develop a particular type of
  logarithmic divergence at the sets where spatial infinity touches
  null infinity. A regularity condition on the initial data can be
  deduced from the analysis of some transport equations on the
  cylinder at spatial infinity. It is given in terms of the linearised
  version of the Cotton tensor and symmetrised higher order
  derivatives, and it ensures that the solutions of the transport
  equations extend analytically to the sets where spatial infinity
  touches null infinity. It is later shown that this regularity
  condition together with the requirement of some particular degree of
  tangential smoothness ensures logarithm-free expansions of the time
  development of the linearised gravitational field close to spatial
  and null infinities.
\end{abstract}

\section{Introduction}

In \cite{Fri98a} an analysis of the behaviour of the gravitational
field close to null infinity and spatial infinity has been given. To
this end, a new representation of spatial infinity was introduced. In
this representation spatial infinity is depicted as a cylinder, as
opposed to the standard representation of it as a point (see e.g.
\cite{Wal84}). This representation has among other things, the
following important feature: it allows to formulate an initial value
problem in which the data and the equations are regular; furthermore,
spacelike infinity and null infinity have a finite representation with
their structure and location known \emph{a priori}. The aforementioned
analysis shows that a certain type of logarithmic divergences arise at
the sets where spatial infinity ``touches'' null infinity (which we
will denote by $I^+$ and $I^-$) unless a certain regularity condition
is satisfied by the initial data. The sources of these logarithmic
divergences can be ultimately traced back to the fact that some of the
evolution equations degenerate at the sets $I^\pm$ (the symbol of the
system looses rank). This precludes the direct application of standard
techniques of partial differential equations if one wishes to push the
solutions of the field equations all the way up to null infinity. This
kind of degeneracies is a peculiarity of not only the gravitational
field, it is also shared by linear massless fields. In order to shed
some light on the nature and consequences of these degeneracies, here
we will look at a simpler situation. Namely, the linearised
gravitational field (spin 2 zero-rest-mass field) propagating on a
Minkowski background. The final product of our analysis will be
isolation of a series of requirements one needs to impose in order to
obtain logarithmic-free expansions close to null and spatial infinity.
One expects, in principle, that it is possible to extend this
discussion to the gravitational field.

\section{Minkowski spacetime close to null and spatial infinity.}

Our point of departure is the usual representation of Minkowski
spacetime $\widetilde{\Mink}$ in Cartesian coordinates:
\[
\widetilde{g}=\widetilde{\eta}_{\mu\nu}dy^\mu dy^\nu,
\]
where $\widetilde{\eta}_{\mu\nu}=\mbox{diag}(1,-1,-1,-1)$. We are
interested in analysing the geometry of this spacetime close to both
null and spatial infinities. Intuitively one expects this region of
spacetime to contain the domain $D=\{y_\mu y^\mu<0\}$. Let us start
by considering the inversion in $D$ given by,
\[
x^\mu=-\frac{y^\mu}{y^\lambda y_\lambda}, \quad
y^\mu=-\frac{x^\mu}{x^\lambda x_\lambda}.
\]
This inversion clearly maps $D$ onto itself. Observe how a point that
is far from the origin of the $y^\mu$ coordinates seems to be close to
the $x^\mu$ origin. In this sense, the $x^\mu$ coordinates convey the
notion of ``lying at infinity''. A simple calculation shows that:
\[
\widetilde{g}=\frac{1}{(x^\lambda
  x_\lambda)^2}\widetilde{\eta}_{\mu\nu}dx^\mu dx^\nu.
\]
Let now $\rho$ be the standard radial coordinate associated with the
spatial coordinates $x^{a}$, $a=1,2,3$. Then $x^\lambda
x_\lambda=(x^0)^2-\rho^2$. Finally, the introduction of a new
coordinate $\tau$ ($|\tau|<1$) via $x^0=\rho\tau$ yields the Minkowski
metric in the form,
\[
\widetilde{g}=\frac{1}{\rho^4(1-\tau^2)^2}\left(d(\rho\tau)^2-d\rho^2-\rho^2d\sigma^2\right),
\]
where $d\sigma^2$ is the standard line element of the unit 2-sphere in
spherical coordinates. The latter line element suggests the
introduction of two different conformal factors. Namely,
\begin{eqnarray*}
&&\Xi=\rho^2(1-\tau^2), \\
&&\Theta=\rho(1-\tau^2).
\end{eqnarray*}
By means of these two conformal factors one
obtains two different extensions 
$\Mink_\Xi$ and $\Mink_{\Theta}$ of the original Minkowski
spacetime near spatial infinity. The choice of the first conformal factor $\Xi$ yields the standard
representation of spatial infinity as a point. Consider the
conformally rescaled metric:
\begin{eqnarray}
\label{g_Xi}
&&g_\Xi=\Xi^2\widetilde{g}=d(\rho\tau)^2-d\rho^2-\rho^2d\sigma^2,  \\
&&\phantom{g_\Omega}=(dx^0)^2 -(dx^1)^2-(dx^2)^2-(dx^3)^2. \nonumber
\end{eqnarray}
Here and all throughout this chapter tilded quantities will always
refer to quantities in the \emph{physical} (i.e. not conformally
rescaled spacetime). It can be shown that all radial spacelike
geodesics in $\widetilde{\Mink}$ map to spacelike curves (not
necessarily geodesics) in $\Mink_\Xi$ with $\{\rho=0\}$ as endpoint.
From (\ref{g_Xi}) we can see that the set of points for which $\rho=0$
is in fact a point, which we denote by $i^0$. On the other hand,
consider the rescaled metric:
\[
g_\Theta=\Theta^2\widetilde{g}=\frac{1}{\rho^2}\left(d(\rho\tau)^2-d\rho^2-\rho^2d\sigma^2\right).
\]
It can also be checked that all radial spacelike geodesics in
$\widetilde{\Mink}$ map to spacelike curves with endpoints at
$\{\rho=0\}$. Now, the metric seems to be singular at $\rho=0$, but
introducing a new coordinate $r=-log(\rho)$ one gets:
\[
g_{\Theta}=d\tau^2-2\tau d\tau dr -(1-\tau)^2dr^2-d\sigma^2.
\]
Observe that $g_{\Theta}|_{\rho=0}=d\tau^2-d\sigma^2$, thus the set
$\{\rho=0\}$ has now the topology of a cylinder with $S^2$ as
spatial sections. We define $I=\{|\tau|<1, \rho=0\}$,
$I^+=\{\tau=1,\rho=0\}$, $I^-=\{\tau=-1,\rho=0\}$, $I^0=\{\tau=0,
\rho=0\}$. 

Let $u,v$ be the null coordinates given by 
\[
u=\rho(1+\tau), \quad v=\rho(1-\tau).
\]
The curves given by $u=u'$, $u'$ a constant and fixed angular
coordinates are null geodesics for which it can be checked that map to
outgoing (future oriented) null geodesics in $\widetilde{\Mink}$.
Similarly, the null geodesics $v=v'$, with $v'$ constant, and fixed
angular coordinates map to incoming (past oriented) null geodesics in
$\widetilde{\Mink}$. Thus the set $\scri^+=\{\tau=1, \rho\neq0\}$ from
which the geodesics with $u=\mbox{constant}$ start corresponds to
future null infinity. Similarly, the set $\scri^-=\{\tau=-1, \rho\neq
0\}$ from which the geodesics with $v=\mbox{constant}$ emanate
corresponds to past null infinity. The sets $I^\pm$ where spatial
infinity ``touches''null infinity are in a particular sense (to be
discussed later) special. Therefore they are considered neither
belonging to the spatial infinity cylinder $I$ nor to $\scri^\pm$.
Notice that the conformal factor $\Theta$ vanishes on $\scri^\pm$, $I$
and $I^\pm$.

\begin{figure}[t]
\centering
\begin{tabular}{cc}
\includegraphics[width=.4\textwidth]{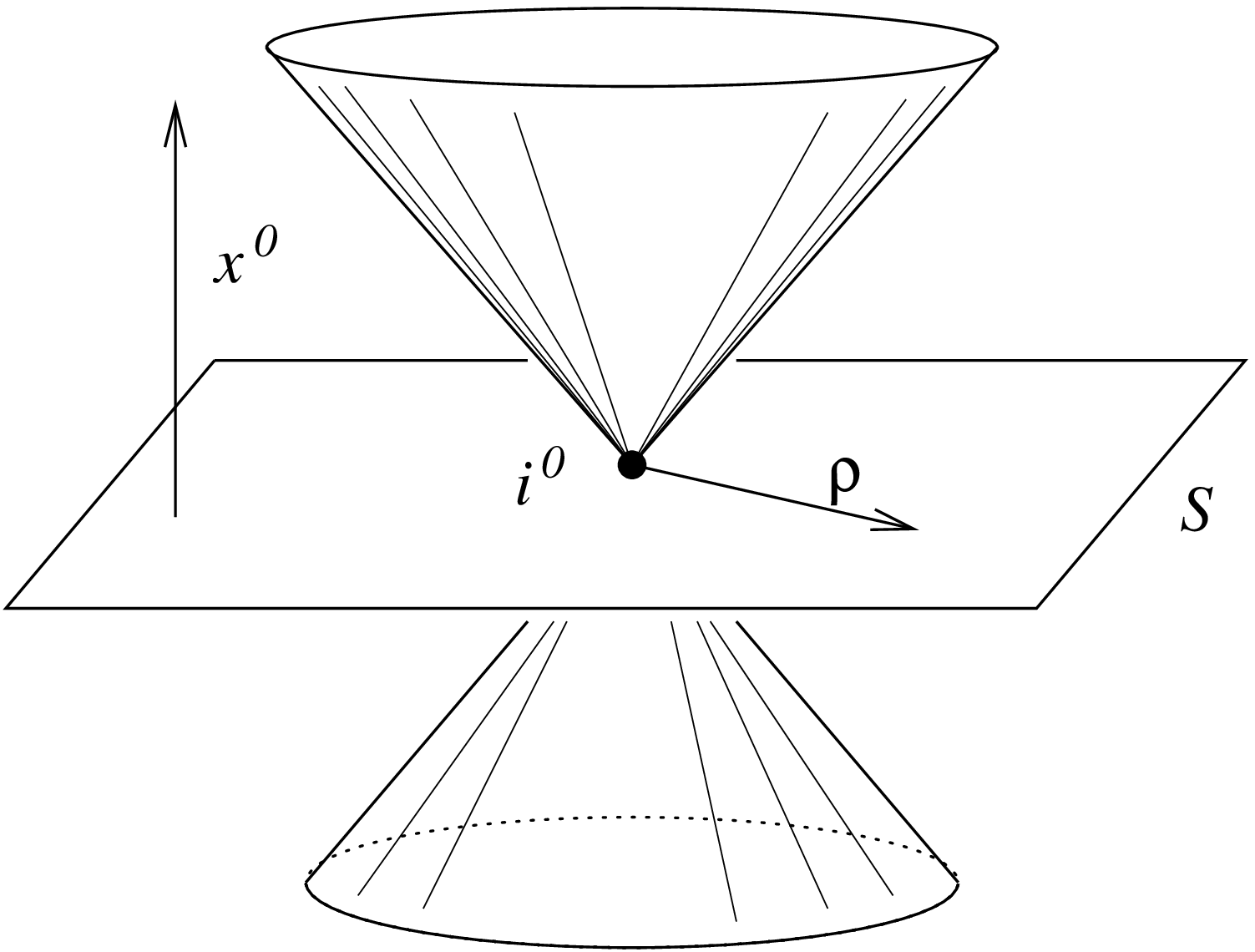} &
\includegraphics[width=.5\textwidth]{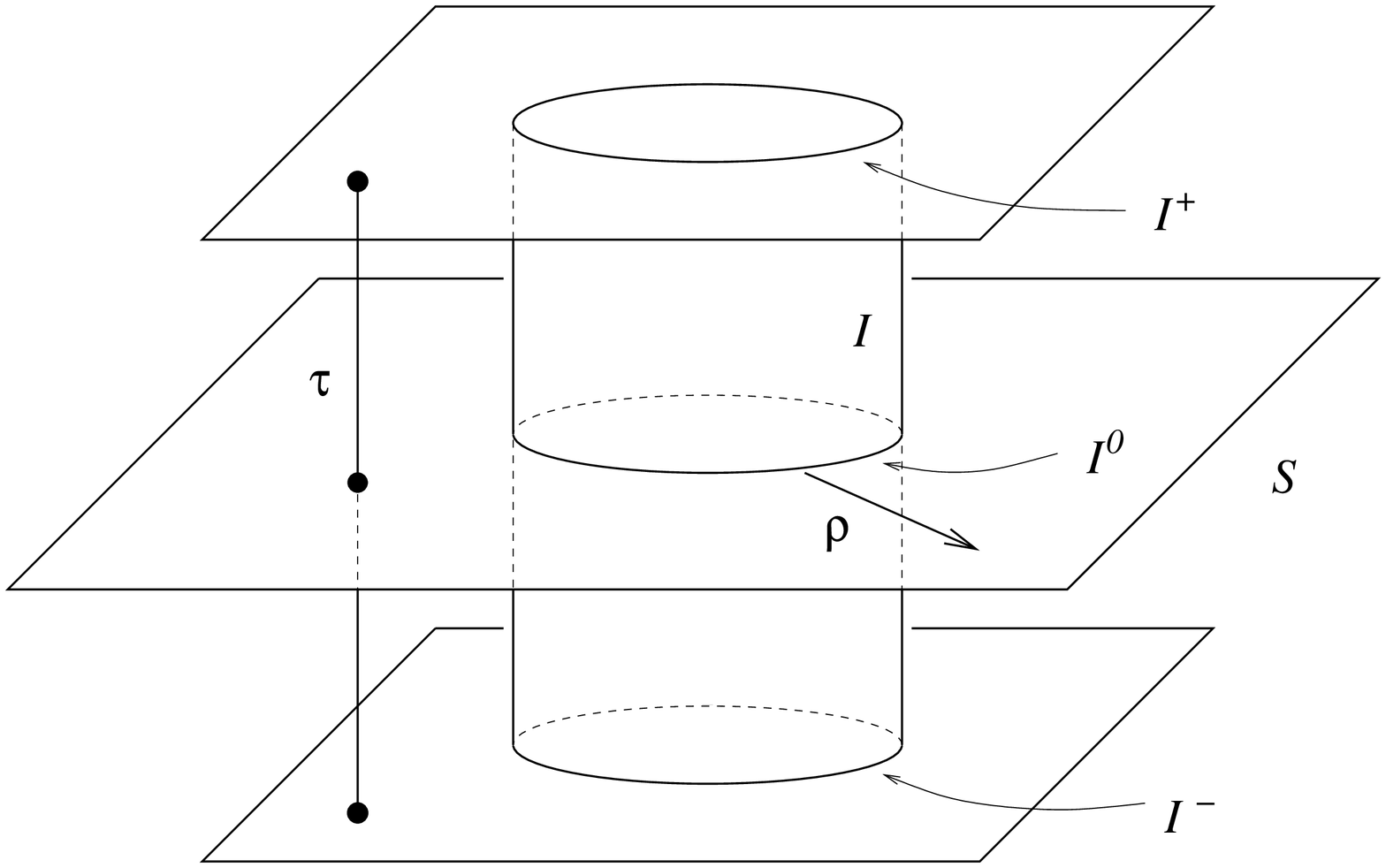}
\end{tabular}
\put(-25,50){$\scri^+$}
\put(-25,-30){$\scri^-$}
\put(-215,30){$\scri^+$}
\put(-215,-30){$\scri^-$}
\caption[]{Spacetime close to spatial and null infinities: to the left
  the standard representation of spatial infinity as a point $i^0$; to
  the right the representation where spatial infinity is enviewed as a
  cylinder.}
\label{characteristic}
\end{figure}

Most of the discussion here presented will be concerned with the
extension $\Mink_\Theta$ of Minkowski spacetime produced by the
conformal factor $\Theta=\rho(1-\tau^2)$.  Using the null geodesics
discussed in the previous paragraph one can construct an adapted null
(NP) tetrad such that the real vector $l$ is tangent along the
incoming null geodesics, and the vector $n$ points along the outgoing
null geodesics \footnote{The use of the vectors $l$ and $n$ here is
  reverse with respect to what it is the standard notation
  \cite{NewPen62,PenRin84}. The vector $l$ is generally chosen to lie
  along outgoing null geodesics. This is done to agree with the
  spinorial notation used in \cite{Fri98a}.}.  The complex vectors $m$
and $\overline{m}$ are then usually chosen so that they span the
tangent space of the two dimensional spheres
$S_{\rho',\tau'}=\{\rho=\rho', \tau=\tau'\}$, with $\rho'$, $\tau'$
given. This construction has the problem that $m$ should vanish at
least in a point on $S_{\rho',\tau'}$. For technical reasons it will
be therefore convenient to coordinatise the 2-spheres $S_{\rho,\tau}$
by means of unitary representations of $SU(2,\Complex)$. This can be
done as the unit sphere $S^2$ can be identified with
$SU(2,\Complex)/U(1)$ which is obtained from $S^3\simeq
SU(2,\Complex)$ via the Hopf map. Any real analytic function $f$ on
$SU(2,\Complex)$ admits an expansion of the form,
\[
f=\sum_{m=0}^\infty \sum_{j=0}^m \sum_{k=0}^m
f_{m,k,j}T_{m\phantom{k}j}^{\phantom{m}k}, 
\]
with complex coefficients $f_{m,k,j}$. The functions
$T_{m\phantom{k}j}^{\phantom{m}k}$ $m=0,1,2,\ldots$, $j,k=0,\ldots,m$
can be related to the spherical harmonics $Y_{l,n}$ \cite{FriKan00}.
The function $f$ will be said to have \emph{spin-weight} $s$ if its
expansion in terms of the $T_{m\phantom{k}j}^{\phantom{m}k}$ functions
is of the form,
\[
f=\sum_{q=|s|}^\infty\sum_{k=0}^{2q} f_{q,k}
T_{2q\phantom{k}q-s}^{\phantom{2q}k}.
\]
Important for later discussion will be the fact that any $C^\infty$
function $\phi=\phi(\rho,t)$ on $\Real_0^+\times SU(2,\Complex)$ such
that for fixed $\rho$ it is of spin-weight $s$, with $s$ independent
of $\rho$ has a \emph{normal
  expansion} of the form,
\[
\phi=\sum_{p\geq|s|}\sum_{q=|s|}^p\sum_{k=0}^{2q}\phi_{p;q,k}T_{2q\phantom{k}q-s}^{\phantom{2q}k}\rho^p,
\]
with $\phi_{p;q,k}$ complex. This last result will be extensively used.
For further details on the $T_{m\phantom{k}j}^{\phantom{m}k}$
functions, the reader is referred to \cite{Fri86a}.

\section{Linearised gravity in the F-gauge}

We will describe the linearised gravitational by means of a spin-2
zero-rest-mass field $\phi_{abcd}$ satisfying,
\[
\nabla^{aa'}\phi_{abcd}=0.
\]
Using the null tetrad described in the previous section, one can
recover the $(D,\overline{\delta})$
and $(\Delta,\delta)$ equations of the NP formalism (see
e.g. \cite{NewPen62,PenRin84}) \footnote{Here again, in order to agree
  with reference \cite{Fri98a} our notation is reversed with respect
  to the standard one of the NP formalism. Our $\phi_0$, $\phi_1$,
  $\phi_2$, $\phi_3$ and $\phi_4$ correspond respectively to the
  $\phi_4$, $\phi_3$, $\phi_2$, $\phi_1$ and $\phi_0$ of the standard
  NP notation.}:
\begin{eqnarray}
&&(1+\tau)\partial_\tau\phi_n-\rho\partial_\rho\phi_n+X_+\phi_{n+1}=(n-2)\phi_n,
  \label{eqn:n}\\
&&(1-\tau)\partial_\tau\phi_{n+1}+\rho\partial_\rho\phi_{n+1}+X_-\phi_n=(n-1)\phi_{n+1}
\label{eqn:nplus1}
\end{eqnarray}
where $n=0,1,2,3$. The coefficients $\phi_n$ can be shown to have
spin-weight $2-n$. The operators $X_+$ and $X_-$ are complex linear
combinations of left invariant vector fields on $SU(2,\Complex)$, and
can be related to the $\eth$ and $\overline{\eth}$ operators of Newman
\& Penrose \cite{FriKan00}. The
NP equations (\ref{eqn:n}) and (\ref{eqn:nplus1}) imply a set of five
\emph{propagation equations},
\begin{eqnarray}
&&(1+\tau)\partial_\tau \phi_0 -\rho\partial_\rho \phi_0 + X_+\phi_1=-2\phi_0,  \label{p0} \\
&&\partial_\tau\phi_1+\textstyle\frac{1}{2}X_-\phi_0 +\frac{1}{2}X_+\phi_2=-\phi_1, \label{p1} \\
&&\partial_\tau\phi_2+\textstyle\frac{1}{2}X_-\phi_1 +\frac{1}{2}X_+\phi_3=0, \label{p2} \\
&&\partial_\tau\phi_3+\textstyle\frac{1}{2}X_-\phi_2 +\frac{1}{2}X_+\phi_4=\phi_3,  \label{p3} \\
&&(1-\tau)\partial_\tau\phi_4 +\rho\partial_\rho \phi_4 + X_-\phi_3=2\phi_4, \label{p4}
\end{eqnarray}
and a set of three \emph{constraint equations},
\begin{eqnarray}
\tau\partial_\tau\phi_1-\rho\partial_\rho\phi_1+\textstyle\frac{1}{2}X_+\phi_2-\frac{1}{2}X_-\phi_0=0, \label{c1} \\
\tau\partial_\tau\phi_2-\rho\partial_\rho\phi_2+\textstyle\frac{1}{2}X_+\phi_3-\frac{1}{2}X_-\phi_1=0, \label{c2} \\
\tau\partial_\tau\phi_3-\rho\partial_\rho\phi_3+\textstyle\frac{1}{2}X_+\phi_4-\frac{1}{2}X_-\phi_2=0. \label{c3}
\end{eqnarray}
This latter set of equations gives rise to equations on the initial
hypersurface $S=\{\tau=0\}$ which correspond to the constraints coming
from the linearised Bianchi identities.

Using the equations (\ref{p0})-(\ref{p4}) one obtains a symmetric
hyperbolic system \cite{CouHil62,Joh91} by simply multiplying
(\ref{p1})-(\ref{p3}) by a factor of 2. The resulting system is of the
form:
\begin{equation}
\label{sh_system}
A^0\partial_\tau\Phi+A^1\partial_\rho\Phi+A^+X_+\Phi+A^-X_-\Phi+B\Phi=0,
\end{equation}
where $A^\mu$ $\mu=0,1,2,3$ and $B$ are $5\times 5$ matrices.  It can
be more concisely written in the form of
$(A^\mu)_{\alpha\beta}\partial_\mu\phi_\beta+B_\alpha=0$.  Crucial for
our discussion will be to realize that the matrix
$A^0=\mbox{diag}(1+\tau,2,2,2,1-\tau)$ looses rank at $\tau=\pm1$.
This \emph{degeneracy} will be the source of most our problems as this
fact precludes the direct use of the standard theory of symmetric
hyperbolic systems.

Let Q be the $5\times 5$ matrix with components given by,
\[
(Q)_{\alpha\beta}=(A^\mu)_{\alpha\beta}\partial_\mu\phi.
\]
Then the characteristics of the system (\ref{sh_system}) are the
hypersurfaces $\phi=k$, with $k$ a real constant, and such that the
scalar field $\phi$ satisfies,
\[
\det Q=0.
\]
It can be readily checked that the hypersurfaces $u=k$ and $v=k'$, with
$u$ and $v$ defined as in the preceeding section are characteristics
of the symmetric hyperbolic system (\ref{sh_system}). In particular,
the hypersurfaces defined by $v=0$ and $u=0$ with $\rho\neq 0$
correspond to $\scri^+$ and $\scri^-$ respectively. Hence, $\scri^+$
and $\scri^-$ are both characteristic hypersurfaces of the system
(\ref{sh_system}). The evaluation of the NP equations
(\ref{eqn:nplus1}) on $\scri^+$ gives rise to \emph{transport
  equations} which enable us to calculate the value of the components
$\phi_1,\ldots,\phi_4$ on $\scri^+$ from a knowledge of the
\emph{radiation field} $\phi_0$ on $\scri^+$. Similarly, using
equations (\ref{eqn:n}) one obtains a corresponding set of transport
equations on $\scri^-$ by means of which it is possible to calculate
the values of $\phi_0,\ldots,\phi_3$ from a knowledge of $\phi_4$ on
$\scri^-$.

A further look to the system (\ref{sh_system}) reveals that at
$\rho=0$ the whole system reduces to transport equations which enable
us to calculate the value of $\phi_n$ ($n=0,\ldots,4$) from
their value at the initial hypersurface $S$. Hence, we say that $I$ is
a total characteristic of (\ref{sh_system}). As a consequence,
no boundary data can be prescribed on $I$.

\subsection{Initial data for linearised gravity}

That the spin 2 zero-rest-mass field $\phi_{abcd}$ can be used to
describe the linearised gravitational field can be seen as
follows. Consider a family $\widetilde{g}_{\mu\nu}(\lambda)$ of
spacetimes depending in a $C^1$ fashion on the parameter $\lambda$,
and such that at $\lambda=0$ one obtains Minkowski spacetime. Therefore,
\[
\widetilde{g}_{\mu\nu}=\widetilde{\eta}_{\mu\nu}+\lambda\widetilde{h}_{\mu\nu}+\O(\lambda^2).
\]
The symmetric rank 2 tensor $\widetilde{h}_{\mu\nu}$ describes the
first order deviation from flatness of the metric
$\widetilde{g}_{\mu\nu}$. The computation of the curvature to first
order in $\lambda$ yields \cite{PenRin84},
\[
\widetilde{K}_{\mu\nu\lambda\rho}=2\widetilde{\nabla}_{[\mu}\widetilde{\nabla}_{|[\nu}\widetilde{h}_{\lambda]|\rho]},
\]
where $\widetilde{\nabla}_\mu$ is the covariant derivative of the
flat spacetime. Then the linearised field equations take the form:
\begin{equation}
\widetilde{K}_{\mu\nu\lambda}^{\phantom{\mu\nu\lambda}\rho}=0.
\end{equation} 
The tensor $\widetilde{K}_{\mu\nu\lambda\rho}$ is trace-free and
possesses the same symmetries of the Riemann tensor. It can be
described in spinorial language by a totally symmetric spinor
$\widetilde{\phi}_{abcd}$ satisfying the spin-2 zero-rest-mass field
equation:
\begin{equation}
\label{zrmeqn_1}
\widetilde{\nabla}^{aa'}\widetilde{\phi}_{abcd}=0.
\end{equation}
Equation (\ref{zrmeqn_1}) is a sufficient condition for the tensor
$\widetilde{K}_{\mu\nu\lambda\rho}$ to be derivable (locally) from
some symmetric tensor $\widetilde{h}_{\mu\nu}$ \cite{SacBer58}. If the
spinorial field $\widetilde{\phi}_{abcd}$ is set to transform as
$\phi_{abcd}=\Omega^{-1}\widetilde{\phi}_{abcd}$ then the equation
(\ref{zrmeqn_1}) is conformally covariant.

Let us assume that the family of metrics
$\widetilde{g}_{\mu\nu}(\lambda)$ arise from the Einstein evolution of
a corresponding 1-parameter family of asymptotically flat, time
symmetric initial data given by the 3-dimensional metric,
\[
{}^{(3)}\widetilde{g}_{\alpha\beta}=\widetilde{\delta}_{\alpha\beta}+\lambda\;\;{}^{(3)}\widetilde{h}_{\alpha\beta}+\O(\lambda^2).
\]
It will be assumed that ${}^{(3)}\widetilde{g}_{\alpha\beta}$
satisfies the time symmetric vacuum constraint equations. Then, the
symmetric tensor ${}^{(3)}\widetilde{h}_{\alpha\beta}$ is a solution
of the corresponding linearised vacuum constraint equations. In the
present discussion we will consider 1-parameter families of 3-metrics
${}^{(3)}\widetilde{g}_{\alpha\beta}$ such that their suitably conformally
rescaled counterparts ${}^{(3)}{g}_{\alpha\beta}$ extend analytically near
$i$, the infinity of the initial Cauchy
hypersurface. For conceptual reasons it is convenient to distinguish
$i$ from $i^0$, the spatial infinity of the whole spacetime. It can be
proved \cite{Fri98a} that under this assumption the
components of the conformally rescaled Weyl spinor of the family of
metrics $g_{\mu\nu}(\lambda)$ in the unphysical spacetime evaluated on
the initial hypersurface $S$ are of the form,
\[
\psi_n(\lambda)|_S=\sum_{p=|2-n|}^\infty\psi_n^p(\lambda)|_S\rho^n,
\]
where
\[
\psi_n^p(\lambda)|_S=\sum_{q=|2-n|}^p\;\sum_{k=0}^{2q}c_{n,p;q,k}(\lambda)T_{2q\phantom{k}q-2+n}^{\phantom{2q}k}
\]
with complex coefficients $c_{n,p;q,k}$ $n=0,1,2,3$. Now, upon
linearisation one obtains a similar behaviour for the components of
the spinorial field $\phi_{abcd}$. Namely,
\begin{equation}
\label{phi_initial_hypersurface}
\phi_n|_{\tau=0}=\sum_{p=|2-n|}^\infty\phi_n^p|_{\tau=0}\rho^n,
\end{equation}
where
\[
\phi_n^p|_{\tau=0}=\sum_{q=|2-n|}^p\sum_{k=0}^{2q}d_{n,p;q,k}T_{2q\phantom{k}q-2+n}^{\phantom{2q}k}.
\]
The components of the Weyl spinor $\psi_n$ satisfy the constraints
coming from the Bianchi identities. Hence, the components $\phi_n$
satisfy automatically their linearised version ---equations
(\ref{c1})-(\ref{c3}) on $S=\{\tau=0\}$. 

Later considerations will make use of a particular spinorial object:
the Cotton spinor (sometimes also called Bach spinor).  The Cotton
spinor $B_{abcd}$ is the 3-dimensional analogue of the Weyl spinor
\cite{Fri88}. It locally characterises conformally flat 3-metrics in
the sense that it vanishes identically if and only if the 3-metric is
locally conformally flat. One can construct its corresponding
linearised version $b_{abcd}$. Linearisation around Minkowski yields
the following expression,
\[
b_{abcd}=2D_{e(a}\Omega\phi_{bcd)}^{\phantom{bcd)}e}+\Omega
D_{e(a}\phi_{bcd)}^{\phantom{bcd)}e},
\]
where $\Omega=\rho^2$ is the conformal factor of the 3-metric
of the Minkowski initial data, and $D_{ab}$ its corresponding
spinorial covariant derivative.

Inspired in a similar result by Friedrich we present now the following
rather technical result. It will be of much use in later discussions.

\begin{lemma}
\label{expansions:lemma}
The coefficients $d_{j,p;p,k}$ of the spin-2 zero-rest-mass
(\ref{phi_initial_hypersurface}) field satisfy the antisymmetry condition
\begin{equation}
\label{antisymmetry}
d_{0,p;p,k}|_{\tau=0}=-d_{4,p;p,k}|_{\tau=0}, \qquad
p=0,1,\ldots,\;\;k=0,\ldots 2p.
\end{equation}
Furthermore,
\[
d_{j,p;p,k}|_{\tau=0}=0, \qquad p=0,\ldots,s^*,\;\;k=0,\dots,2p, \;\;j=0,\ldots,4.
\]
if and only if
\[
D_{(a_qb_q}\cdots D_{a_1b_1}b_{abcd)}(i)=0, \qquad q=0,\ldots,s^*.
\]
\end{lemma}

The proof follows from direct linearisation of theorem 4.1 in
\cite{Fri98a}.

\section{A regularity condition at spatial infinity}

We now proceed to carry out an analysis of the transport equations one
obtains upon evaluation of the field equations (\ref{p0})-(\ref{p4})
and (\ref{c1})-(\ref{c3}). Consider the equations (\ref{p0})-(\ref{p4}) and
(\ref{c1})-(\ref{c3}). Differentiating them formally with respect to
$\rho$ and evaluating at the cylinder at $\rho=0$ one gets:
\begin{eqnarray}
&&(1+\tau)\partial_\tau \phi^p_0 + X_+\phi_1^p-(p-2)\phi_0=0,  \label{cylp0} \\
&&\partial_\tau\phi^p_1+\textstyle\frac{1}{2}(X_-\phi^p_0 +X_+\phi^p_2)+\phi_1=0, \label{cylp1} \\
&&\partial_\tau\phi^p_2+\textstyle\frac{1}{2}(X_-\phi^p_1 +X_+\phi^p_3)=0, \label{cylp2} \\
&&\partial_\tau\phi^p_3+\textstyle\frac{1}{2}(X_-\phi^p_2 +X_+\phi^p_4)-\phi^p_3=0,  \label{cylp3} \\
&&(1-\tau)\partial_\tau\phi^p_4 + X_-\phi^p_3+(p-2)\phi^p_4=0, \label{cylp4}
\end{eqnarray}
and
\begin{eqnarray}
\tau\partial_\tau\phi^p_1+\textstyle\frac{1}{2}(X_+\phi^p_2-X_-\phi^p_0)-p\phi_1^p=0, \label{cylc1} \\
\tau\partial_\tau\phi^p_2+\textstyle\frac{1}{2}(X_+\phi^p_3-X_-\phi^p_1)-p\phi_2^p=0, \label{cylc2} \\
\tau\partial_\tau\phi^p_3+\textstyle\frac{1}{2}(X_+\phi^p_4-X_-\phi^p_2)-p\phi_3^p=0, \label{cylc3}
\end{eqnarray}
where we have set $\phi^p_n=\partial_\rho^{(p)}\phi_n|_{\rho=0}$. One can
expand these coefficients using the functions
$T_{k\phantom{l}m}^{\phantom{k}l}$ 
in the form:
\begin{equation}
\label{phi_np}
\phi^p_n=\sum_{q=|2-n|}^p\sum_{k=0}^{2q}
a_{n,p;q,k}T_{2q\phantom{k}q-2+n}^{\phantom{2q}k}.
\end{equation}
The coefficients $a_{n,p;q,k}$ are, in principle, complex functions of
$\tau$.  Using these expansions and the equations
(\ref{cylp1})-(\ref{cylp3}) and (\ref{cylc1})-(\ref{cylc3}), one can
calculate the coefficients $a_{1,p;q,k}$, $a_{2,p;q,k}$, $a_{3,p;q,k}$
from a knowledge of the $a_{0,p;q,k}$ and $a_{4,p;q,k}$ via a first
order algebraic linear system. Some more algebra leads to,
\begin{eqnarray}
&&\!\!\!\!\!\!\!\!\!\!\!\!(1-\tau^2)\ddot{a}_{0,p;q,k}+(4+2(p-1)\tau)\dot{a}_{0,p;q,k}+(q+p)(q-p+1)a_{0,p;q,k}=0,
\label{jacobi1} \\
&&\!\!\!\!\!\!\!\!\!\!\!\!(1-\tau^2)\ddot{a}_{4,p;q,k}+(-4+2(p-1)\tau)\dot{a}_{4,p;q,k}+(q+p)(q-p+1)a_{4,p;q,k}=0,
\label{jacobi2}
\end{eqnarray}
for $p\geq 2$, $2\leq q \leq p$, the overdot denoting differentiation
with respect to $\tau$. The equations (\ref{jacobi1}) and
(\ref{jacobi2}) are examples of Jacobi equations. A canonical
parametrisation for this class of ordinary differential equations is:
\begin{equation}
D_{(n,\alpha,\beta)}a\equiv
(1-\tau^2)a^{\prime\prime}+\{\beta-\alpha-(\alpha+\beta+2\tau)\}a^\prime+n(n+\alpha+\beta+1)a=0.
\label{canonic_jacobi}
\end{equation}
In our case the parameters are given by: $\alpha_0=\beta_4=-p-2$,
$\beta_0=\alpha_4=-p+2$, $n_1=p+q$, and $n_2=p-q-1$. Regular solutions
for equations (\ref{jacobi1}) and (\ref{jacobi2}) exist for $q \neq
p$, and are given quite concisely in terms of Jacobi polynomials \cite{Sze78}.
Namely,
\begin{eqnarray}
&&a_{0,p;q,k}=C_1P_{p-q-1}^{(-p-2,-p+2)}(\tau)+C_2\left(\frac{1-\tau}{2}\right)^{p+2}P_{q-2}^{(p+2,-p+2)}(\tau),\label{solution_0}
\\
&&a_{4,p;q,k}=D_1P_{p-q-1}^{(-p+2,-p-2)}(\tau)+D_2\left(\frac{1-\tau}{2}\right)^{p-2}P_{q+2}^{(p-2,-p-2)}(\tau),\label{solution_4}
\end{eqnarray}
where $C_1$, $C_2$, $D_1$ and $D_2$ are constants which can be
determined from the initial data at $\tau=0$. In the case
$q=p$, the use of some identities of the Jacobi
equation (\ref{canonic_jacobi}) leads to:
\begin{eqnarray}
&&\!\!\!\!\!\!\!\!\!\!\!\!\!a_{0,p;p,k}=\left(\frac{1-\tau}{2}\right)^{p+2}\left(\frac{1+\tau}{2}\right)^{p-2}\left(E_0+E_1\int_0^\tau\frac{ds}{(1+s)^{p-1}(1-s)^{p+3}}\right)
\label{sing_sol_0} \\
&&\!\!\!\!\!\!\!\!\!\!\!\!\!a_{4,p;p,k}=\left(\frac{1-\tau}{2}\right)^{p-2}\left(\frac{1+\tau}{2}\right)^{p+2}\left(F_0+F_1\int_0^\tau\frac{ds}{(1+s)^{p+3}(1-s)^{p-1}}\right).
\label{sing_sol_4}
\end{eqnarray}
The use of Taylor expansions in the integrals shows that they give
rise to logarithmic terms. More precisely, 
recalling that:
\begin{eqnarray*}
&&\int_0^\tau\frac{ds}{(1\pm s)^{p-1}(1\mp
  s)^{p+3}}=A_{*}\ln(1-\tau)+\frac{A_{p\pm 2}}{(1-\tau)^{p\pm
    2}}+\cdots+\frac{A_{1}}{(1-\tau)}+A_0
\nonumber \\
&&\phantom{\int_0^\tau\frac{ds}{(1+s)^{p-1}(1-s)^{p+3}}=}
+B_{*}\ln(1+\tau)+\frac{B_{p\mp 2}}{(1+\tau)^{p\mp 2}}+\cdots+\frac{B_{1}}{(1+\tau)},
\end{eqnarray*}
where the $A$'s and $B$'s are some constants. From the latter
expansions one can conclude that the only non-regular coefficients in
$a_{0,p;q,k}$ and $a_{4,p;q,k}$ in the $q=p$ case are of logarithmic
nature. Analytic solutions arise if and only if
$a_{0,p;p,m}(0)=a_{4,p;p,m}(0)$. That is only the case if the
constants $E_1$ and $F_1$ in equations (\ref{sing_sol_0}) and
(\ref{sing_sol_4}) are both zero. Now, using the antisymmetry
condition (\ref{antisymmetry}) of lemma \ref{expansions:lemma} one
learns that in fact $a_{0,p;p,m}(0)=a_{4,p;p,m}(0)=0$. It is not hard
to see that if this is the case, then $a_{j,p;p,m}(0)=0$ for
$j=0,\dots,4$. Finally, using the second part of lemma
\ref{expansions:lemma} one can relate this behaviour of the initial
data to the vanishing of the linearised Cotton tensor and its
symmetrised derivatives on $i$. The results of the previous discussion
are summarised in the following theorem.

\begin{theorem}
\label{theorem:regularity_condition}
The solutions of the transport equations on $I$ corresponding to the
system (\ref{p0})-(\ref{p4}) and (\ref{c1})-(\ref{c3}) extend
analytically to $I^\pm$ if and only if the \emph{regularity condition}
\begin{equation}
\label{regcond}
D_{(a_sb_s\cdots}D_{a_1b_1}b_{abcd)}(i)=0, \quad s=0,1,\dots \;\; , 
\end{equation}
holds.
\end{theorem}
A peculiarity of the logarithms appearing in the solutions of the transport
equations is that they only occur at the highest spherical harmonics
sector at each order. This will be of importance later when
discussion the asymptotic expansions of the field $\phi_{abcd}$ close
to $\scri^+$. 

The aforediscussed solutions of the transport equations allow us to
obtain \emph{normal expansions} of the form,
\begin{equation}
\label{normal_expansions}
\phi_n=\sum_{p\geq|2-n|}\phi_n^p \rho^p,
\end{equation}
with the coefficients $\phi_n^p$ given by (\ref{phi_np}). Now, it is
of interest to see how the form of these normal expansions and the
regularity condition (\ref{regcond}) reflect on the structure of the
asymptotic expansions close to null infinity. In order to do this, one
has essentially to reshuffle the normal expansions. So far, our
discussion has applied to both future and past null infinities. The
forthcoming analysis will without loss of generality be focused on
$\scri^+$. Nevertheless, a totally analogous treatment can be
performed for $\scri^-$. 

Putting together the results from the analysis of the transport
equations (\ref{cylp0})-(\ref{cylp4}) and (\ref{cylc1})-(\ref{cylc3})
one obtains normal expansions for the coefficients $\phi_0$ and
$\phi_4$ which are of the form,
\begin{eqnarray*}
&&\phi_0=\sum_{p=2}\left(S[0](1-\tau)+a_*^p\ln(1-\tau)P[p+2,2p](1-\tau)\right)\rho^p,
  \\
&&\phi_4=\sum_{p=2}\left(S[0](1-\tau)+b_*^p\ln(1-\tau)P[p-2,2p](1-\tau)\right)\rho^p.
\end{eqnarray*}
By $S[n](x)$ it will be understood a generic infinite series in
$x$ starting with $x^n$. Similarly, $P[n_1,n_2](x)$ denotes generic
polynomials in $x$ of order $n_2$ and whose lowest order term is
$x^{n_1}$. The coefficients $a_*^p$, $b_*^p$ and those in the
series/polynomials are given in terms of the functions
$T_{m\phantom{k}j}^{\phantom{m}k}$. Matters of convergence of the
normal expansions will be addressed in the next section. Now, a
careful reshuffling in order to obtain expansions in $(1-\tau)$ reveal
that,
\begin{eqnarray*}
&&\!\!\!\!\!\!\!\phi_0=\sum_{p=0}^3S[2](\rho)(1-\tau)^p+\!\!\sum_{p\geq
  4}\left(S[2](\rho)+ 
 \!\!\!\! \sum_{s=[t/2]}^{t-2}a^s_*\rho^s \ln(1-\tau)\right)(1-\tau)^p, 
\\
&&\!\!\!\!\!\!\!\phi_4=\sum_{p\geq 0}\left(S[2](\rho)+ 
 \!\!\!\! \sum_{s=[t/2]}^{t+2}b^s_*\rho^s \ln(1-\tau)\right)(1-\tau)^p,
\end{eqnarray*}
where $[t/2]=t/2$ if $t$ is even, and $[t/2]=(t+1)/2$ if $t$ is
odd. Notice that the component $\phi_4$ happens to be the most
singular one of the field. If the regularity condition (\ref{regcond})
is satisfied up to $s=s_*$, then it is not hard to see that,
\begin{equation}
\label{cond1}
\phi_4=\sum_{k=0}^{s_*+1}\ c_{k}(1-\tau)^k + d_{s_*+1}(1-\tau)^{s_*+1}
\ln(1-\tau)+\cdots,
\end{equation}
where the coefficients $c_k$ are $C^\infty$ functions of $\rho$ and the
angular variables, and 
\begin{equation}
\label{cond2}
d_{s_*+1}=\rho^{s_*+3}\sum_{m=0}^{2s_*+6}D_{s_*+1,m}
T_{2s_*+6\phantom{m}s_*+5}^{\phantom{2s_*+6}m},
\end{equation}
with $D_{s_*+1,m}$ complex constants. Thus, the logarithmic term in the
expansion has only dependence in the highest spherical harmonic
sector possible at this order, i.e.  $q=s_*+3$. Finally, if the
regularity condition holds also for $s=s_*+1$ then $d_{s_*+1}=0$.

\section{Polyhomogeneous expansions}

\subsection{A substraction argument}

So far, nothing has been said about the convergence of the normal
expansions (\ref{normal_expansions}) calculated in the previous
section. We will now discuss how this can be done. As before, we write
$\Phi=(\phi_0,\phi_1,\phi_2,\phi_3,\phi_4)$.  Let,
\[
\Phi_N=\sum_{p=0}^N\frac{1}{p!}\Phi^p\rho^p,
\]
be the $N$ order partial sum. The linear field equations
(\ref{p0})-(\ref{p4}) and (\ref{c1})-(\ref{c3}) are such 
that $\Phi_N$ itself happens to be a solution of the field
equations with truncated (to order $N$) initial data. The regularity
of $\Phi_N$ will depend on whether the initial data satisfies the
regularity condition (\ref{regcond}) to a given order or not. For
example, if the condition is satisfied up to $s=N$, then no
$\ln(1-\tau)$ term will be present in the coefficients $\Phi^p$, and
thus $\Phi_N$ will be $C^\infty$. How could one estimate the
rest? Recall that the symmetric hyperbolic system derived from
equations (\ref{p0})-(\ref{p4}) breaks down precisely at $\tau=1$.
This is rather unfortunate, as we are mainly interested in observing
the behaviour of the solutions on null infinity.

\begin{figure}[t]
\centering
\begin{tabular}{cc}
\includegraphics[width=.5\textwidth]{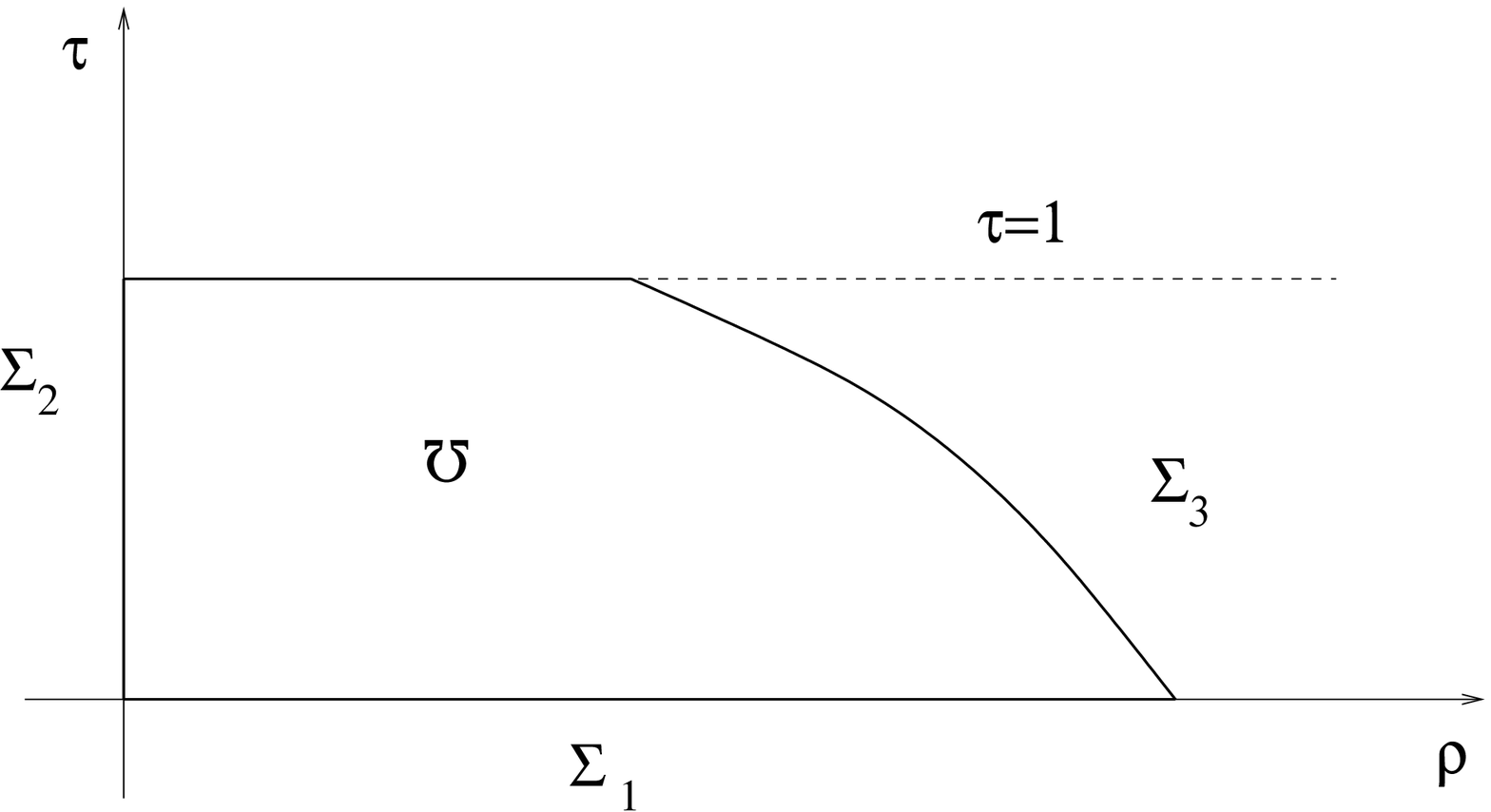} &
\includegraphics[width=.5\textwidth]{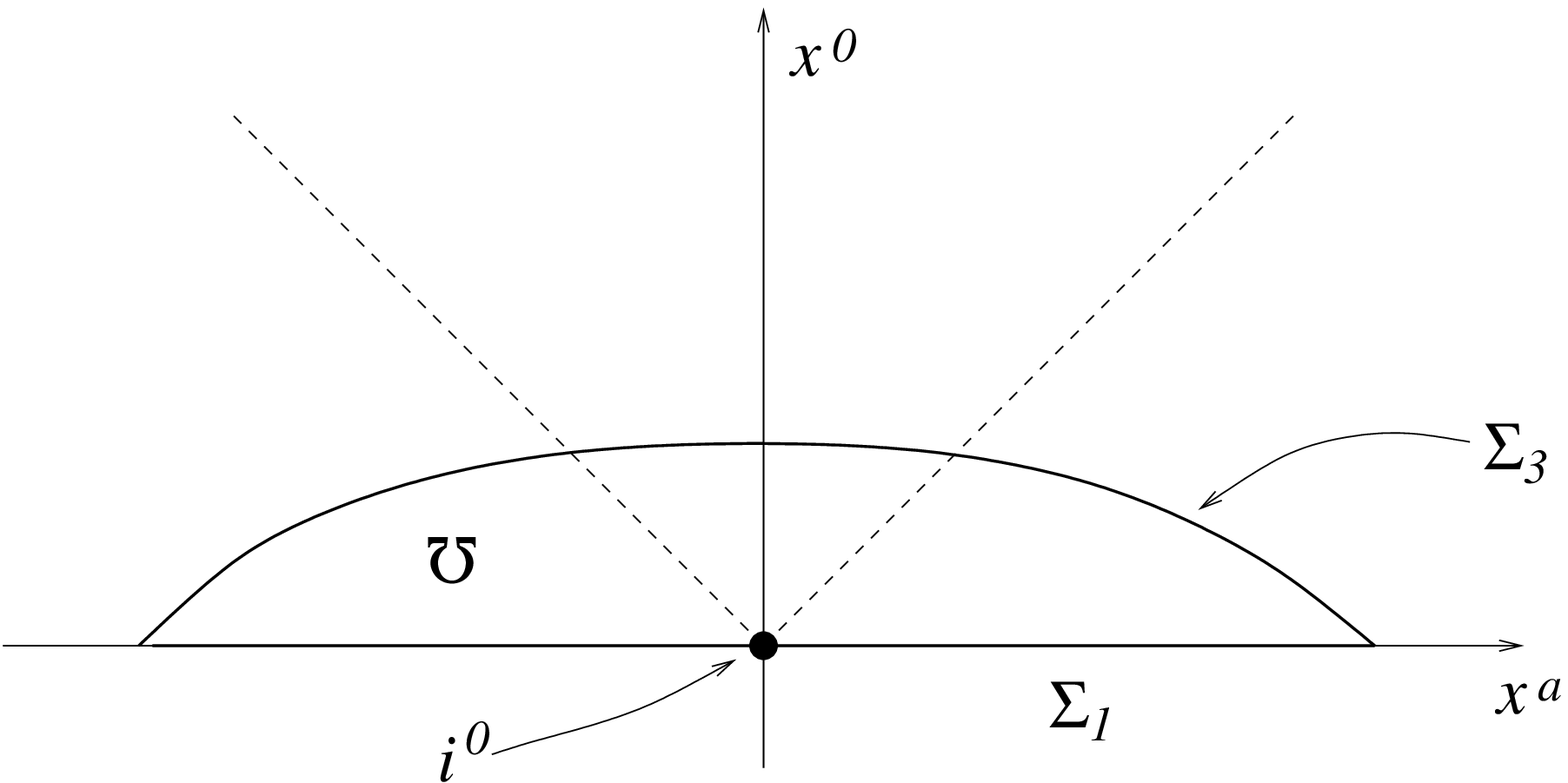}
\end{tabular}
\put(-220,25){$\scri^+$}
\put(-60,25){$\scri^+$}
\caption[]{The region of integration used to obtain the energy estimate
  (\ref{energy_estimate}). The first picture shows the domain of
  integration in the representation where spatial infinity is enviewed
  as a cylinder, and the second in the in the standard representation
  of spatial infinity as a point. Note that $\Sigma_2$ which seems to
  be an extended 2 dimensional surface is in fact a point: $i^0$. The
  surface $\Sigma_3$ is spacelike and almost flat, and may reach and
  go through null infinity.}
\label{integration_region}
\end{figure}

Bounds for the rest can be obtained by considering the conformal
extension of Minkowski, $\Mink_\Xi$, in which spatial infinity is
represented by a point $i^0$. The spin 2 zero-rest-mass field
equations given in spherical coordinates are formally singular at
$\rho=0$ in this representation. Therefore, we resort to Cartesian
coordinates $(x^\mu)$. In this way, the propagation equations read,
\begin{eqnarray*}
&&\partial_0 \phihat_0 -\partial_3 \phihat_0
+\partial_1\phihat_1+i\partial_2\phihat_1=0, \\ 
&&2\partial_0\phihat_1+\partial_1\phihat_2+\partial_1\phihat_0+i\partial_2\phihat_2-i\partial_2\phihat_0=0,
\\ 
&&2\partial_0\phihat_2+\partial_1\phihat_3+\partial_1\phihat_1+i\partial_2\phihat_3-i\partial_2\phihat_1=0,
\\ 
&&2\partial_0\phihat_3+\partial_1\phihat_4+\partial_1\phihat_2+i\partial_2\phihat_4-i\partial_2\phihat_2=0,
\\ 
&&\partial_0 \phihat_4 +\partial_3 \phihat_4 +\partial_1\phihat_3-i\partial_2\phihat_3=0, 
\end{eqnarray*}
 while the constraint equations are given by, 
\begin{eqnarray*}
&&2\partial_3\phihat_1+\partial_1\phihat_0-\partial_1\phihat_2-i\partial_2\phihat_0-i\partial_2\phihat_2=0,
\\ 
&&2\partial_3\phihat_2+\partial_1\phihat_1-\partial_1\phihat_3-i\partial_2\phihat_1-i\partial_2\phihat_3=0,
\\ 
&&2\partial_3\phihat_3+\partial_1\phihat_2-\partial_1\phihat_4-i\partial_2\phihat_2-i\partial_2\phihat_4=0. 
\end{eqnarray*}
The components of the linearised Weyl tensor rescale as follows:
$\phihat_n=\rho^{-1-n}\phi_n$, $n=0,\ldots,4$, where
$\rho^2=(x^1)^2+(x^2)^2+(x^3)^2$.  Thus, the smooth data in
$\Mink_\Theta$ discussed in \S 3.1, and such that $\phi_0=\O(\rho^2)$,
$\phi_1=\O(\rho)$, $\phi_2=\O(1)$, $\phi_3=\O(\rho)$ and
$\phi_4=\O(\rho^2)$, becomes singular in the $\Mink_\Xi$ picture at
$i$ (i.e. $\rho=0$).  However, if one provides initial data with
sufficiently fast decay at $i^0$ then one can obtain energy estimates
of the form,
\begin{equation} \label{energy_estimate}
\|\widehat{\Phi}\|^2_{L^2(\mho)} \leq C
\|\widehat{\Phi}\|^2_{L^2(\Sigma_1)},
\end{equation}
where $\|\phantom{X}\|^2_{L^2(\mho)}$ and
$\|\phantom{X}\|^2_{L^2(\Sigma_1)}$ denote the $L^2$ norms over the
region $\mho$ \footnote{The symbol $\mho$ is to be read ``mho''.} and
the hypersurface $\Sigma_1$ respectively.  Higher order estimates for
the derivatives can be similarly obtained. The domain $\mho$ is as
shown in figure (\ref{integration_region}). Notice how in this way one
can obtain estimates that go up to $\scri^+$ ---and through it.
In this spirit consider, 
\[
\widehat{X}=\widehat{\Phi}-\widehat{\Phi}_N. 
\]
A close look to the rescaling rules will convince us that:
\[
\widehat{X}|_{\tau=0}=\O(\rho^{N-5}).
\]
It can be readily checked that such a decay implies
$\widehat{X}|_{\tau=0}\in H^{N-4}(\Sigma_1)$, where $H^k(\Sigma_1)$
denotes the $k$-th Sobolev space over the domain $\Sigma_1$ .
Standard theory of linear of symmetric hyperbolic systems (see for
example \cite{Tay96}) then guarantees the existence of $\widehat{X}$
such that $\widehat{X}\in C^{N-7}(\mho)$. Hence, we have estimated the
rest of the partial sum $\Phi_N$. If one wants the solution to be of
class $C^k(\mho)$, $k\in\Natural$, then the previous discussion shows
that one needs the regularity condition (\ref{regcond}) to hold up to
order $s=k+7$.

The disadvantage of this approach is that it is hard to extend for the
case of quasilinear equations. Therefore, it will be of little use
when we eventually try to address the full non-linear gravitational
field. We need to devise something else.

\subsection{An investigation of expansions close to null infinity}

The analysis of the solutions of the transport equations on the the
spatial infinity cylinder $I$ show that logarithmic divergences will
generically arise at $I^\pm$ for smooth initial data of the sort
considered in \S 3.1 unless the regularity condition (\ref{regcond})
is satisfied. Accordingly, if the regularity condition is satisfied,
then the solutions to the transport equations extend smoothly through
$I^+$.  This is a hint that the evolution of generic smooth initial
data will be \emph{polyhomogeneous}, i.e. its expansions will contain
$\ln(1-\tau)$ terms. An analysis and a discussion of the properties of
such spacetimes close to null infinity, including the existence of an
intriguing set of conserved quantities analogous to the Newman-Penrose
constants has been given in
\cite{ChrMacSin95,Val98,Val99a,Val99b,Val00a}. The regularity
condition given in theorem \ref{theorem:regularity_condition} ensures
that the solutions of the transport equations
(\ref{cylp0})-(\ref{cylp4}) and (\ref{cylc1})-(\ref{cylc3}) extend
analytically to $I^\pm$. Now, with the discussion on $I^+$ more or
less settled, one would like to study the effects the degeneracy of
the field equations on $I^+$ has one null infinity. In particular, one
would like to know what do the logarithmic terms on $I^+$ imply on
$\scri^+$. How do they propagate? In order to investigate this point,
we resort to some ideas coming from the \emph{asymptotic
  characteristic initial value problem} \cite{Fri81b,Fri81a,Kan96b}.
Consider a future oriented null hypersurface $\mathcal{N}_0$
intersecting $\scri^+$ at $\mathcal{Z}_0$. One can provide
\emph{characteristic initial data} in the following way: $\phi_4$ is
prescribed on $\mathcal{N}_0$, $\phi_0$ on the portion of $\scri^+$
comprised between $\mathcal{Z}_0$ and $I^+$, and $\phi_2$ and $\phi_1$
on $\mathcal{Z}_0$. Theorems providing existence in a neighbourhood
$\mho\subset J^-(\mathcal{Z}_0)$ of $\mathcal{Z}_0$ for the conformal
Einstein equations have been given in \cite{Fri81b,Fri81a,Kan96b}. An
existence theorem for polyhomogeneous Maxwell fields has been given in
\cite{Val00}.
\begin{figure}[t]
\centering
\includegraphics[width=.4\textwidth]{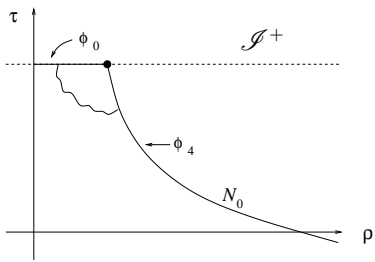}
\put(-50,80){$\scri^+$} 
\caption[]{The asymptotic characteristic initial value problem close to
  spatial infinity.}
\label{figure2}
\end{figure}
Problems with the characteristic approach have mainly to do with the
ample freedom one has in choosing the initial data on $\mathcal{N}_0$.
As an added factor, the theorems by Friedrich and K\'{a}nn\'{a}r given
only local existence. This is somehow a problem, as one would like to
take the field $\phi_{abcd}$ all along $\scri^+$ and evaluate at the
set $I^+$, and then compare with the results obtained from the
transport equations on $I$.

Inspired by the aforementioned characteristic approach, we will make
use of the transport equations obtained on $\scri^+$ arising from the
NP equations (\ref{eqn:nplus1}). In this way once the radiation field
$\phi_0$ is provided, the remaining field components can be obtained
essentially by a mere integration along the generators of null infinity.
Expansions containing $\ln(1-\tau)$ terms appeared in the normal
expansions (\ref{normal_expansions}). However, it is of interest to
consider a more general type of situation in which terms of the form
$\ln^k(1-\tau)$ can arise. So, we will assume that the components
of the spin 2 zero-rest-mass field $\phi_{abcd}$ have asymptotic
expansions near to $\scri^+$ of the form,
\begin{equation}
\label{Ansatz}
\phi_n=\sum_{i\geq0}\sum_{j=0}^{N_i} \phi_n^{(i,j)}(1-\tau)^i\ln^j
(1-\tau),
\end{equation}
$=0,\ldots,4\;$, where the coefficients $\phi_n^{(i,j)}$ contain only
$\rho$ and angular dependence. They are assumed to be smooth at least
in $\{\tau=1, 0<\rho\geq \rho_0\}$ for $\rho_0$ a given non-negative
real number. Some calculus rules will be needed in order to manipulate
the polyhomogeneous expansions (\ref{Ansatz}).

\textbf{Assumption.} \emph{For the expansions (\ref{Ansatz}), it will be
assumed that,}
\begin{eqnarray}
&&\partial_\tau\phi_n=-\sum_{i\geq 1} \sum_{j=0}^{N_i}\left(i\ln^j(1-\tau)+j\ln^{j-1}(1-\tau)\right)\phi_n^{(i,j)}(1-\tau)^{i-1},
\label{taudiff}\\
&&\partial_\rho\phi_n=\sum_{i\geq 0} \sum_{j=0}^{N_i}
\phi_n^{(i,j)'}(1-\tau)^i\ln^j(1-\tau), \label{rhodiff}\\
&&X_{\pm} \phi_n=\sum_{i\geq 0} \sum_{j=0}^{N_i}X_{\pm}
\phi_n^{(i,j)}(1-\tau)^i\ln^j(1-\tau), \label{xdiff}
\end{eqnarray}
\emph{hold, where $'$ denotes differentiation with respect to $\rho$.} 

The above calculus rules can be used together with the transport
equations on $\scri^+$ to obtain equations for the coefficients
$\phi_n^{(i,j)}$ in the Ansatz (\ref{Ansatz}). At each order
$(1-\tau)^p$ the solutions to these equations will be calculated. The
conditions under which the coefficients $\phi_n^{(i,j)}$ can be pushed
down to will then be investigated. The idea is then to compare the
expansions obtained by means of this procedure with those arising from
solving the transport equations on $I$. If an identification of the
two different expansions is possible, then one can analyse the effect
the regularity condition (\ref{regcond}) has on the asymptotic
expansions close to spatial infinity $\scri^+$. Now, if at each order
$(1-\tau)^p$ it is assumed that no $\log^k(1-\tau)$ terms are present
in the lower terms, then it is possible to implement an inductive
argument. The results of our forthcoming argument are summarised in
the following theorem.

\begin{theorem}
Let, 
\begin{itemize}
\item[i)] the components of the spin 2 zero-rest-mass field $\phi_{abcd}$
  have an evolution expansion of the form
\[
\phi_n=\sum_{i\geq 0}\sum_{j=0}^{N_i} \phi_n^{(i,j)}(1-\tau)^i\ln^j
(1-\tau),
\]
$n=0,\dots,4$ with the coefficients $\phi_n^{(i,j)}$ at least $C^1$ in
$\{\tau=1, 0<\rho\leq\rho_0\}$ for $\rho_0$ a non-negative constant;
\item[ii)] the calculus rules (\ref{taudiff})-(\ref{xdiff}) hold;
\item[iii)] $\phi_0^{(0,0)}$ (the radiation field)
  be $C^\infty$ in $\{\tau=1, 0\leq\rho\leq\rho_0 \}\subset\scri^+\cup
  I^+$;
\item[iv)] $\phi_1^{(0,0)}$ and $\phi_3^{(0,0)}$ be bounded in $\{\tau=1, 0\leq\rho\leq\rho_0 \}$;
\item[v)] the initial data $\phi_n|_{\tau=0}$, $n=0,\ldots,4$, be obtained
  by linearisation around flat space of asymptotically flat time
  symmetric initial data satisfying the Einstein vacuum constraints,
  and analytic in a neighbourhood of $i$;
\item[vi)] furthermore, the initial data satisfy the regularity condition
\[
D_{(a_sb_s}\cdots D_{a_1b_1}b_{abcd)}(i)=0 \quad s=0,1,\dots;\]
\item[vii)] $\partial_\rho^{(k)}\phi_4^{(i,0)}$ be continuous in
  $\{\tau=1, 0\leq\rho\leq\rho_0 \}$, for $k=0,\ldots,i+2$ (tangential
  smoothness).
\end{itemize}
Then the expansions of $\phi_n$ are in fact logarithm-free,
that is,
\[
\phi_n=\sum_{i\geq 0}\phi_n^{(k,0)}(1-\tau)^k,
\]
where the coefficients $\phi_n^{(i,0)}$ are $C^\infty$ in $\{\tau=1, 0\leq\rho\leq\rho_0 \}$.
\end {theorem}

\subsubsection{The base step $(1-\tau)^0$.}
In agreement with the Ansatz (\ref{Ansatz}) let the leading terms of
the spin 2 zero-rest-mass field be of the form,
\[
\phi_n=\sum_{j=0}^{N_0}\phi_n^{(0,j)}\ln^j(1-\tau)+\sum_{i\geq 1}\sum_{j=0}^{N_i}\phi_n^{(i,j)}(1-\tau)^{i}\ln^j(1-\tau),
\]
with $n=0\ldots4$. Direct substitution into equations (\ref{eqn:n})
show that,
\[
\phi_0^{(0,j)}=\phi_1^{(0,j)}=\phi_2^{(0,j)}=\phi_3^{(0,j)}=0,
\]
for $j\geq 1$. The coefficients $\phi_0^{(0,0)}$, $\phi_1^{(0,0)}$,
$\phi_2^{(0,0)}$ can be determined from the equations
(\ref{eqn:nplus1}) $n=0,1,2$. The coefficients satisfy,
\begin{eqnarray}
&&\rho\phi_1^{(0,0)\prime}+X_{-}\phi_0^{(0,0)}+\phi_1^{(0,0)}=0, \label{phi1scri} \\
&&\rho\phi_2^{(0,0)\prime}+X_{-}\phi_1^{(0,0)}=0, \label{phi2scri}\\
&&\rho\phi_3^{(0,0)\prime}+X_{-}\phi_2^{(0,0)}-\phi_3^{(0,0)} =0, \label{phi3scri}
\end{eqnarray}
which can in principle be solved if the coefficient
$\phi_0^{(0,0)}=\phi_0|_{\scri^+}$ (the radiation field) is given. It
will be assumed that this is the case, and  moreover, that it is
$C^\infty$.  
Being $\phi_0^{(0,0)}$ smooth, one has that
\begin{equation}
\phi^{(0,0)}_0= \sum_{k\geq2}f_{0,k}^{(0)}\rho^k =\sum_{k\geq 2}\left( \sum_{q=2}^{k}\sum_{m=0}^{2q}F^{(0)}_{0,k;q,m}
T_{2q\phantom{m}q-2}^{\phantom{2q}m}\right)\rho^k, \label{phi0scri}
\end{equation}
with $F^{(0)}_{0,k;q,m}$ complex constants. Using this expansion and
equations (\ref{phi1scri})-(\ref{phi3scri}) plus the further
requirement of $\phi_1$ and $\phi_3$ to be smooth at $I^+$ one finds
that,
\begin{equation}
\label{base_step_orders}
 \phi_1=\O(\rho^2), \quad \phi_2=\O(1), \quad \phi_3=\O(\rho).
\end{equation}
The substitution of the Ansatz for the time development in equation
(\ref{eqn:nplus1}) $n=3$, yields as a result the following hierarchy:
\begin{eqnarray*}
&&-2\phi_4^{(0,N_{0})}+\rho\phi_4^{(0,N_{0})\prime}=0, \nonumber \\
&&-2\phi_4^{(0,N_{0}-1)}-\phi_4^{(0,N_{0})}+\rho\phi_4
^{(0,N_{0}-1)\prime}=0, \nonumber \\
&&\;\;\;\;\;\;\; \vdots \nonumber \\
&&-2\phi_4^{(0,0)}-\phi_4^{(0,1)}+\rho\phi_4^{(0,0)\prime}
+X_{-}\phi_3^{(0,0)}=0. \nonumber
\end{eqnarray*}
It can be solved from top to bottom, yielding:
\begin{eqnarray}
&& \phi_4^{(0,N_{0})}=
\phihat_4^{(0,N_{0})}\rho^{2}, \nonumber \\
&& \phi_4^{(0,N_{0}-1)}=
\phihat_4^{(0,N_{0}-1)}\rho^{2}+ \phihat_4^
{(0,N_{0})}\rho^{2}\ln\rho, \nonumber \\
&&\phantom{\phi_4^{(0,N_{0}-1)}=}\vdots \nonumber \\
&& \phi_4^{(0,0)}=\left[\frac{1}{N_0!}\phihat_4^{(0,N_{0})}\rho^{2}
\ln^{N_{0}}\rho+\cdots+\phihat_4^{(0,0)}\rho^{2}\right] 
-\rho^{2}\int X_{-}\phi_3^{(0,0)}\rho^{-3}d\rho \label{phi40j}
\end{eqnarray}
where the hatted quantities denote functions which appear during the
integration and thus, contain only angular dependence. The coefficients
$\phi_4^{(0,j)}$ can be then ``pushed down'' to $I^+$, but not in
general in such a way that they are smooth at $I^+$. The presence of
terms of the form $\ln^k(1-\tau)$ in the Ansatz for the evolution
yields as a result the presence of $\ln^{N_0-k}\rho$ terms in the
corresponding expansions. This is an indication that if $\ln\rho$ terms
are present in the initial data then these will propagate in the
evolution via some particular $\ln^k(1-\tau)$ terms. Logarithmic terms
of the form $\ln^j\rho$ do not appear in the normal expansions
(\ref{normal_expansions}). Thus, in order to be able of performing an
identification of the expansions on $\scri^+$ and $I$ some
degree of \emph{tangential smoothness} at $I^+$ will be assumed. A
close look to formulae (\ref{phi40j}) shows that one needs
$\phi_4^{(0,0)}$ to have two continuous $\rho$-derivatives at $I^+$. We will denote this by $\phi_4^{(0,0)}\in C^2_\rho(I^+)$. This
requirement now implies that $\phihat_4^{(0,j)}=0$ for $j\geq 2$, and
thus also $\phi_4^{(0,j)}=0$ for $j\geq 2$. Note that the
``integration constant'' $\phihat_4^{(0,1)}$ does not need to vanish
as the $\rho^2\ln\rho$ term in $\phi_4^{(0,0)}$ may be canceled with a
similar one coming from $\rho^{2}\int
X_{-}\phi_3^{(0,0)}\rho^{-3}d\rho$. Indeed, the integral $\rho^{2}\int
X_{-}\phi_3^{(0,0)}\rho^{-3}d\rho$ produces a term of the form
$\rho^2\ln\rho$ from the $\rho^2$ term in $\phi_3^{(0,0)}$. As
discussed previously,
\begin{equation}
\phi_3^{(0,0)}=\sum_{k\geq 1}f_{3,k}^{(0)}\rho^k, \label{phi300}
\end{equation}
where the coefficients $f^{(0)}_{3,k}$ are calculated from the
$f^{(0)}_{0,k}$ of equation (\ref{phi0scri}) by means of equations
(\ref{phi1scri})-(\ref{phi3scri}). Setting
$\phihat_4^{(0,1)}=X_{-}f_{3,2}^{(0)}$ in accordance with our previous
discussion one finds that,
\begin{eqnarray}
&&\phi_4^{(0,1)}=X_{-}f_{3,2}^{(0)}\rho^{2},  \label{phi401}\\
&&\phi_4^{(0,0)}=\phihat_4^{(0,0)}\rho^{2}- \sum_{\substack{k\geq1 \\ k\neq2}} \frac{X_{-}f_{3,k}^{(0)}}{k-2}\rho^k. \label{phi400}
\end{eqnarray}
One expects to be able to deduce the required degree of tangential
smoothness form the smoothness properties of the initial
data.
 
Finally, one is now in the position of performing an identification of
the expansions obtained from the analysis of the transport equations
on the cylinder, and those obtained from the transport equations at
$\scri^+$. As discussed, both expansions contain $\ln(1-\tau)$ terms.
However, the $\ln(1-\tau)$'s in the normal expansions appear only at
the highest spherical sector at each order. So do
those in our asymptotic expansions: from equation (\ref{phi300}) one
has,
\[
f_{3,2}^{(0)}=\sum^2_{q=1}\sum_{m=0}^{2q}F^{(0)}_{3,2;q,m}T_{2q\;\;\;q+1}^{\;\;\;m}.
\] 
And hence, 
\[
X_{-}f_{3,2}^{(0)}=-\sum_{q=1}^2\sum_{m=0}^{2q}F^{(0)}_{3,2;q,m}\beta_{2q,q+2}T_{2q\;\;\;q+2}^{\;\;\;m},
\]
with $\beta_{2q,q+2}=\sqrt{(q+2)(q-1)}$. Thus, $X_{-}$ cancels the
sector given by $q=1$, and as a consequence the only non-zero sector
in (\ref{phi401}) is $q=2$. It is noticed by passing that
$X_-\phi_4^{(0,1)}=0$ whence it follows that (tangential smoothness
assumed) $X_-\phi_4$ is bounded at $I^+$. Hence, the leading term of
the evolution expansion of $\phi_4$ has the same form of the expansion
given in (\ref{cond1}) and (\ref{cond2}). Consequently, if
$b_{abcd}(i)=0$ (regularity condition at order $s=0$) holds, we have
shown that,
\begin{equation}
\label{end_base_step}
\phi_n=\phi_n^{(0,0)}+\sum_{i\geq1}\sum_{j=0}^{N_i}\phi_n^{(i,j)}(1-\tau)^{i}\ln^j(1-\tau),
\end{equation}
$n=0,\dots,4$. That is, there are no logarithms in the leading term
of the expansions.

\subsubsection{The step $(1-\tau)^1$.}
The base induction step is somehow exceptional. In order to understand
better the situation for the general case it is convenient to take a
look to what happens at the order $(1-\tau)^1$. To begin with, let us
assume that the analysis of base step $(1-\tau)^0$ has been carried
out, and no $\ln(1-\tau)$ are present at that order. Consequently, the
expansion (\ref{end_base_step}) holds.

Using equations (\ref{eqn:nplus1}) $n=0,1,2$ and a similar approach to
the one used in the base step, one readily finds that:
\[
\phi_0^{(1,j)}=\phi_1^{(1,j)}=\phi_2^{(1,j)}=\phi_3^{(1,j)}=0,
\]
for $j\geq 1$. This is a direct consequence of the fact that there are
no logarithms at order $(1-\tau)^0$. Furthermore,
\begin{eqnarray}
&& \phi_0^{(1,0)}=\frac{1}{2}\left[ 2\phi_0^{(0,0)}-\rho\phi_0
^{(0,0)\prime}+X_{+}\phi_1^{(0,0)}\right],  \label{phi010}\\
&& \phi_1^{(1,0)}=\frac{1}{2}\left[ \phi_1^{(0,0)}-\rho\phi_1
^{(0,0)\prime}+X_{+}\phi_2^{(0,0)}\right],  \label{phi110} \\ 
&& \phi_2^{(1,0)}=\frac{1}{2}\left[-\rho\phi_2^{(0,0)\prime}
+X_{+}\phi_3^{(0,0)}\right],                \label{phi210}\\  
&& \phi_3^{(1,0)}=\frac{1}{2}\left[ -\phi_3^{(0,0)}-\rho\phi_3
^{(0,0)\prime}+X_{+}\phi_4^{(0,0)}\right].  \label{phi310}
\end{eqnarray}
Thus, one can calculate the coefficients $\phi_0^{(1,0)}$,
$\phi_1^{(1,0)}$, $\phi_2^{(1,0)}$ and $\phi_3^{(1,0)}$ directly from
a knowledge of the coefficients at order $p=0$. Using
(\ref{base_step_orders}) in equations (\ref{phi010})-(\ref{phi310})
one concludes that,
\[
\phi_n^{(1,0)}=\sum_{k\geq |2-n|}f^{(1)}_{n,k}\rho^k.
\]
The coefficients $f^{(1)}_{n,k}$ given in
terms of the coefficients $f^{(0)}_{n,k}$ of the previous order. As in
the base step, the logarithmic dependence comes from equation
(\ref{eqn:nplus1}) with $n=3$. From it one obtains,
\begin{eqnarray*}
&&-3\phi_4^{(1,N_{1})}+\rho\phi_4^{(1,N_{1})\prime}=0, \nonumber \\
&&-3\phi_4^{(1,N_{1}-1)}-\phi_4^{(1,N_{1})}+\rho\phi_4
^{(1,N_{1}-1)\prime}=0, \nonumber \\
&&\;\;\;\;\;\;\; \vdots \nonumber \\
&&-3\phi_4^{(1,0)}-\phi_4^{(1,1)}+\rho\phi_4^{(1,0)\prime}
+X_{-}\phi_3^{(1,0)}=0, 
\end{eqnarray*}
which again is solved from top to bottom so that,
\begin{eqnarray*}
&&\!\!\!\!\! \phi_4^{(1,N_{1})}=\phihat_4^{(1,N_{1})}\rho^{3}, \nonumber \\
&&\!\!\!\!\! \phi_4^{(1,N_{1}-1)}=\phihat_4^{(1,N_{1}-1)}\rho^{3}+ \phihat_4^{(1,N_{1})}\rho^{3}\ln\rho, 
\nonumber \\
&&\!\!\!\!\!\phantom{\phi_4^{(1,N_{1}-1)}}\vdots \nonumber \\
&&\!\!\!\!\!\phi_4^{(1,0)}=
\left[\frac{1}{N_1!}\phihat_4^{(1,N_{1})}\rho^{3}\ln^{N_{1}}\rho+\cdots
+\phihat_4^{(1,0)}\rho^{3}\right] 
-\rho^{3}\int X_{-}\phi_3^{(1,0)}\rho^{-4}d\rho. 
\end{eqnarray*}
Whence one observes one more time the appearance of $\ln^k \rho$
dependence associated with the $\ln^{N_1-k}(1-\tau)$ in the Ansatz for
the evolution.  The expression $\rho^{3}\int
X_{-}\phi_3^{(1,0)}\rho^{-4}d\rho$ can only account for the
cancellation of the $\rho^3\ln\rho$ term in $\phi_4^{(1,0)}$ via the
$\rho^3$ term in $\phi_3^{(1,0)}$.  The coefficient $\phi_3^{(1,0)}$
itself contains no logarithmic dependence, as a consequence of
equation (\ref{phi310}) and the smoothness of the coefficients
calculated at the base step $p=0$. Following the spirit
of previous discussions, we require $\phi_4^{(1,0)}\in C^3_\rho(I^+)$.
Therefore $\phihat_4^{(1,j)}=0$ for $j=2,\dots,N_1$, and furthermore
$\phi_4^{(1,j)}=0$ for $j=2,\dots,N_1$. So, one is left with only two
non-zero coefficients. Namely,
\begin{eqnarray*}
&&\phi_4^{(1,1)}=\phihat_4^{(1,1)}\rho^3,\\  
&&\phi_4^{(1,0)}=\phihat_4^{(1,1)}\rho^3\ln\rho+\phihat_4^{(1,0)}\rho^{3} 
-\rho^{3}\int X_{-}\phi_3^{(1,0)}\rho^{-4}d\rho 
\end{eqnarray*}
Now, writing 
\begin{equation}
\phi_3^{(1,0)}=\sum_{k\geq 1}f^{(1)}_{3,k}\rho^k, \label{phi310:a}
\end{equation}
one has to set
\begin{equation}
\phihat_4^{(1,1)}=X_-f^{(1)}_{3,3}, \label{phihat411}
\end{equation}
in order to eliminate the remaining logarithmic term in
$\phi_4^{(1,0)}$. 

In a similar way to what happened in the base step, one can further
show that the logarithmic dependence is only found at the highest
spherical harmonic sector. However, in this case and also in the
general case, the analysis is a bit more elaborate. From equations
(\ref{phi401}) and (\ref{phi310}) one finds that:
\[
\phi_3^{(1,0)}={\textstyle\frac{1}{2}}\left[\sum_{k\geq
    1}(-1-k)f_{3,k}^{(0)}\rho^k+X_+\phihat_4^{(0,0)}\rho^2-\sum_{k\geq
    3}{\textstyle\frac{X_{+}X_{-}f_{3,k}^{(0)}}{k-2}} \rho^k \right].
\]
The spin weight of $\phi_3$ is $-1$, and accordingly,
\[
f_{3,k}^{(0)}=\sum_{q=1}^k\sum_{m=0}^{2q}F^{(0)}_{3,k;q,m}T_{2q\;\;\;q+1}
^{\;\;\;m}.
\]
Thus, at the end of the day one has the following rather
complicate-looking expression:
\begin{eqnarray*}
&&\phi_3^{(1,0)}=\sum_{k\geq 2} f_{3,k}^{(1)}\rho^k \nonumber \\
&&\phantom{\phi_3^{(1,0)}}={\textstyle \frac{1}{2}}(X_+\phihat_4^{(0,0)}-f^{(0)}_{3,2})\rho^2
\nonumber \\
&&\phantom{\phi_3^{(1,0)}=}{\textstyle 
+\frac{1}{2}}\!\!\!\sum_{\substack{k\geq1\\k\neq2}}\sum_{q=1}^k
\sum_{m=0}^{2q}\left(\textstyle{\frac{(q+2)(q-1)}{k-2}-(k+1)}\right)F^{(0)}_{3,k;q,m}
T_{2q\;\;\;q+1}^{\;\;\;m}\rho^k.
\end{eqnarray*} 
The term $(q+2)(q-1)-(k+1)(k-2)$ vanishes whenever $q=k-1$. Hence, if 
one writes
\[
f_{3,k}^{(1)}=\sum_{q=1}^{k}\sum_{m=0}^{2q}F_{3,k;q,m}^{(1)}T_{2q\;\;\;q+1}
^{\;\;\;m},
\]
one has $F^{(1)}_{3,k;k-1,m}=0$ for $k=1,2,\ldots\;\;$. Therefore, the
$\rho^3$ term in $X_{-}\phi_3^{(1,0)}$ contains only $q=3$ and $q=1$
spherical harmonic dependence, i.e.
\[
f^{(1)}_{3,3}=\sum_{m=0}^2
F^{(1)}_{3,3;1,m}T_{2\phantom{m}2}^{\phantom{2}m}+\sum_{m=0}^6F^{(1)}_{3,3;3,m}T_{6\phantom{m}4}^{\phantom{6}m}.
\]
Therefore, using (\ref{phihat411}) one finally arrives to:
\begin{equation}
\phi_4^{(1,1)}=-4\rho^3\sum_{m=0}^6F^{(1)}_{3,3;3,m}T_{6\phantom{m}5}^{\phantom{6}m}.
\end{equation}
Again, the expansions have the same form of (\ref{cond1}) and
(\ref{cond2}) with $s_*=0$. If the regularity condition is satisfied
up to $s=1$, i.e. if $b_{abcd}(i)=0$ and $D_{(a_1b_1}b_{abcd)}(i)=0$
then
\[
\phi_n=\phi_n^{(0,0)}+\phi_n^{(1,0)}(1-\tau)+\sum_{i\geq2}\sum_{j=0}^{N_i}\phi_n^{(i,j)}(1-\tau)^{i}\ln^j(1-\tau).
\]

\subsubsection{The general step $(1-\tau)^{p+1}$.}
The procedures of the general step are lengthier but nevertheless of a
similar nature to those of the analysis of the $(1-\tau)^1$ terms. We
begin by assuming that a similar analysis to that one carried for
$(1-\tau)^1$ has been carried up to $(1-\tau)^p$ inclusive.
Accordingly, the components of the field are assumed to have an
expansion of the form,
\begin{equation}
\phi_n=\sum_{k=0}^p\phi_n^{(k,0)}(1-\tau)^k + \sum_{k\geq p+1} \sum_{j=0}^{N_{k}} 
\phi_n^{(k,j)}(1-\tau)^k\ln^j(1-\tau) \label{orderp}.
\end{equation}
The coefficients $\phi_n^{(k,0)}$ with $k\leq p$ are $C^\infty$, for
they have been constructed out of the radiation field $\phi_0^{(0,0)}$
in a way that preserves the smoothness. Hence, 
\[
\phi_n^{(p,k)}=\sum_{k\geq|2-n|}f_{n,k}\rho^k,
\]
where the coefficient $f_{n,k}$ contains all the angular dependence.
In particular, for the component $\phi_3$ they are such
that,
\[
f_{3,k}^{(p)}=\sum_{q=1}^{k}\sum_{m=0}^{2q}F^{(p)}_{3,k;q,m}
T_{2q\phantom{m}q+1}^{\phantom{2q}m},
\]
with coefficients $F^{(p)}_{3,k;q,m}$ such that
\begin{equation}
F^{(p)}_{3,k;k-1,m}=F^{(p)}_{3,k;k-2,m}=\dots=F^{(p)}_{3,k;k-p,m}=0, \label{indhyp}
\end{equation}
as suggested by the analysis of the $(1-\tau)^0$ and $(1-\tau)^1$
cases. Substitution of the Ansatz (\ref{orderp}) in equations
(\ref{eqn:n}) with $n=0,1,2,3$ yields a hierarchy of $N_{p+1}+1$
equations. A similar analysis to that performed at order $(1-\tau)$
readily shows that
\[
\phi_0^{(p+1,j)}=\phi_1^{(p+1,j)}=\phi_2^{(p+1,j)}=\phi_3^{(p+1,j)}=0,
\]
for $1\leq j \leq N_{p+1}$. One also obtains a set of 4 recurrence
relations which allows to calculate the coefficients
$\phi_n^{(p+1,0)}$ $n=0,1,2,3$ from (in principle known) lower order
terms. In future discussions only the expression for
$\phi_3^{(p+1,0)}$ will be needed,
\begin{equation}
\phi_3^{(p+1,0)}={\textstyle \frac{1}{2(p+1)}}\left[ (p-1)\phi_3^{(p,0)}-\rho\phi_3
^{(p,0)\prime}+X_{+}\phi_4^{(p,0)}\right]. \label{phi3pplus10}
\end{equation}
The coefficients $\phi_4^{(p+1,j)}$ are on the other hand calculated
using equation (\ref{eqn:nplus1}) with $n=3$. Similar computations to
those described in the $(1-\tau)^0$ and $(1-\tau)^1$ steps lead to:
\begin{eqnarray}\label{hierarchy_phi4pplus1}
&& \phi_4^{(p+1,N_{p+1})}=\phihat_4^{(p+1,N_{p+1})}\rho^{p+3}, \nonumber \\
&& \phi_4^{(p+1,N_{p+1}-1)}=\textstyle \phihat_4^{(p+1,N_{p+1}-1)}\rho^{p+3}+ \phihat_4^
{(p+1,N_{p+1})}\rho^{p+3}\ln\rho, \nonumber \\
&&\phantom{\phi_4^{(p+1,N_{p+1})}}\vdots \nonumber \\
&&\phi_4^{(p+1,0)}={\textstyle\frac{1}{N_p!}}\phihat_4^{(p+1,N_{p+1})}\rho^{p+3}
\ln^{N_{p+1}}\rho+\cdots+\phihat_4^{(p+1,0)}\rho^{p+3}
\nonumber \\
&&\phantom{\phi_4^{(p+1,0)}=XX}-\rho^{p+3}\int X_{-}\phi_3^{(p+1,0)}\rho^{-4-p}d\rho. 
\end{eqnarray}
Again, the term $\rho^{p+3}\int X_{-}\phi_3^{(p+1,0)}\rho^{-4-p}d\rho$
in the hierarchy (\ref{hierarchy_phi4pplus1}) can be only used to
cancel out the $\phihat_4^{(p+1,1)}\rho^{p+3}\ln\rho$ term in the
expression for the coefficient $\phi_4^{(p+1,0)}$. Thus, in order to
perform an identification at $I^+$, some tangential smoothness will be
required. More precisely, we will demand $\phi_4^{(p+1,0)}\in
C_\rho^{p+3}(I^+)$. This assumption implies that
$\phihat_4^{(p+1,N_{p+1})}=\ldots=\phihat_4^{(p+1,2)}=0$, with the
further consequence of yielding
$\phi_4^{(p+1,N_{p+1})}=\ldots=\phi_4^{(p+1,2)}=0$. Hence as before,
we are left with only two non-zero coefficients:
\begin{eqnarray}
&&\phi_4^{(p+1,1)}=\phihat_4^{(p+1,1)}\rho^{p+3}, \\ 
&&\phi_4^{(p+1,0)}=\phihat_4^{(p+1,1)}\rho^{p+3}\ln \rho +\phihat_4^{(p+1,0)}
\rho^{p+3} \nonumber \\
&&\phantom{XXXXXXXXXXXX}-\rho^{p+3}\int X_{-}\phi_3^{(p+1,0)}\rho^{-4-p}d\rho. \label{phi4pplus10}
\end{eqnarray}
The coefficient $\phi_3^{(p+1,0)}$ is smooth, as it is constructed
from the $\phi_n^{(p,0)}$\,'s 
using equation (\ref{phi3pplus10}). Consequently, 
\begin{equation}
\phi_3^{(p+1,0)}=\sum_{k\geq 1}f_{3,k}^{(p+1)}\rho^k, \label{phi3pplus10:a}
\end{equation}
with the coefficients $f_{3,k}^{(p+1)}$ containing only angular
dependence. Hence, one sees that:
\[
\rho^{p+3}\int X_{-}\phi_3^{(p+1,0)}\rho^{-4-p}d\rho=X_{-}f_{3,p+3}^{(p+1)}
\rho^{p+3}\ln\rho+\sum_{\substack{k\geq1 \\k\neq p+3}} \textstyle{\frac{X_{-}f_{3,k}^{(p+1)}}{k-p-3}}
\rho^k. 
\]
In accordance with our previous discussion one sets: 
\[
\phihat_4^{(p+1,1)}= X_{-}f_{3,p+3}^{(p+1)}.
\]
In this way we have accounted for the remaining logarithm in
$\phi_4^{(p+1,0)}$ ---see equation (\ref{phi4pplus10}). Therefore, at
the end of the day one has that:
\begin{eqnarray*}
&&\phi_4^{(p+1,1)}=X_{-}f_{3,p+3}^{(p+1)}\rho^{p+3}, \\
&&\phi_4^{(p+1,0)}=\phihat_4^{(p+1,0)}\rho^{p+3}- \sum_{\substack{k\geq2\\k\neq p+3}} \textstyle{\frac{X_{-}f_{3,k}^{(p+1)}}{k-p-3}}
\rho^k.
\end{eqnarray*}

The component $\phi_4$ is the most irregular one of the field, and
thus, the $\ln(1-\tau)$ terms will appear firstly there. Carrying out
the previous discussion to the following orders, one finds that the
first $\ln(1-\tau)$ terms in $\phi_3$, $\phi_2$, $\phi_1$ and $\phi_0$
appear at orders $(1-\tau)^{p+2}$, $(1-\tau)^{p+3}$, $(1-\tau)^{p+4}$
and $(1-\tau)^{p+5}$ respectively.

Finally, in order to make use of the regularity condition
(\ref{regcond}) it is again necessary to identify the $\ln(1-\tau)$'s
appearing in our expansions with those coming from the analysis of the
transport equations on the cylinder $I$. Using equations
(\ref{phi3pplus10}) and (\ref{phi3pplus10:a}) one finds
\[
\phi_3^{(p+1,0)}=\frac{1}{p+1}\left[G_{p+2}\rho^{p+2}+\sum_{k\geq 1}(p-1-k)f_{3,k}^{(p)}\rho^k-\sum_{\substack{k\geq
    1\\ k\neq p+2}}{\textstyle\frac{X_{+}X_{-}f_{3,k}^{(p)}}{k-p-2}} \rho^k \right].
\]
One can make explicit the spherical harmonics dependence:
\begin{eqnarray*}
&&\phi_3^{(p+1,0)}={\textstyle\frac{1}{p+1}}\Biggl[G_{p+2}\rho^{p+2} \\
&&\phantom{XXXXX}+\sum_{\substack{k\geq1 \\ k\neq p+2}}\sum_{q=1}^k
\sum_{m=0}^{2q}\left( {\textstyle\frac{(q+2)(q-1)}{k-p-2}+(p-k-1)}\right)F^{(p)}_{3,k;q,m}
T_{2q\;\;\;q+1}^{\;\;\;m}\rho^k\Biggr], \label{big}
\end{eqnarray*}
where the coefficients $F^{(p)}_{3,k;q,m}$ satisfy (\ref{indhyp}). The
expression $(k-p+1)(k-p-2)-(q+2)(q-1)$ is zero whenever $q=p-k$ and/or
$q=k-p-1$. Hence, writing
\[
f_{3,k}^{(p+1)}=\sum_{q=1}^{k}\sum_{m=0}^{2q}F_{3,k;q,m}^{(p+1)}T_{2q\;\;\;q+1}
^{\;\;\;m},
\]
then one finds that
\[
F^{(p+1)}_{3,k;k-p-1,m}=0.
\]
Furthermore, using the induction hypothesis (\ref{indhyp}) in equation 
(\ref{big}) one finds that
\[
F^{(p+1)}_{3,k;k-1,m}=F^{(p+1)}_{3,k;k-2,m}=\ldots= F^{(p+1)}_{3,k;k-p,m}=
F^{(p+1)}_{k;k-p-1,m}=0. 
\]

In particular if $k=p+3$ then $X_{-}f_{p+3}^{(p+1)} \rho^{p+3}\ln\rho$
will only contain spherical harmonic dependence at the $q=p+3$ sector.
The other possibly remaining sector, the $q=1$ one, is annihilated by
the $X_{-}$ derivative. Accordingly, it has been proved that
\[
\phi_4^{(p+1,1)}=-\beta_{2p+6,p+5}\rho^{p+3}\sum_{m=0}^{2p+6}
F_{3,p+3;p+3,m}^{(p+1)}T_{2p+6\phantom{m}p+5}^{\phantom{2p+5}m}.
\]
Hence, the expansions we have so far obtained are similar to those
given in (\ref{cond1}) and (\ref{cond2}). Finally, if the regularity
condition (\ref{regcond}) holds up to order $p+1$ then one has,
\[
\phi_n=\sum_{k=0}^{p+1}\phi_n^{(k,0)}(1-\tau)^k + \sum_{k\geq p+2} \sum_{j=0}^{N_{k}} \phi_n^{(k,j)}(1-\tau)^k\ln^j(1-\tau).
\]

\subsubsection{Concluding remarks.}

Some final remarks come now into place. One would expect to be able to
deduce the smoothness of $\phi_0$ on null infinity, the boundedness of
$\phi_1$ and $\phi_3$ and the tangential smoothness of $\phi_4$ from
the smoothness of the initial data. This would require, in principle,
the implementation of some energy estimates which would allow us to
reach null infinity (see the discussion in \S 3). This is at the time
of writing still an open problem. The present analysis is nevertheless
valuable in the sense that it focuses our attention on the
facts/properties one should be able to  prove, and their interconnections. It
should be, in principle, possible to undertake an analogue of the
above discussion for the non-linear gravitational field. The analysis
of the linearised gravitational field has shown us the way to
lead. These matters will be the subject of future work. 

\subsubsection*{Acknowledgements.}
I would like to thank H. Friedrich for suggesting me this research
topic, for numerous discussions and criticisms which lead to the
substantial improvement of this work.


\begin{thebibliography}{10}

\bibitem{ChrMacSin95}
P.~T. Chru\'{s}ciel, M.~A.~H. Mac{Callum}, \& D.~B. Singleton,
\newblock {\em Gravitational waves in general relativity {X}{I}{V}. {Bondi}
  expansions and the ``polyhomogeneity'' of $\scri$},
\newblock Phil. Trans. Roy. Soc. Lond. A {\bf 350}, 113 (1995).

\bibitem{CouHil62}
R.~Courant \& D.~Hilbert,
\newblock {\em Methods of Mathematical Physics}, volume~{I}{I},
\newblock John Wiley \& Sons, 1962.

\bibitem{Fri81b}
H.~Friedrich,
\newblock {\em The asymptotic characteristic initial value problem for
  {Einstein}'s vacuum field equations as an initial value problem for a
  first-order quasilinear symmetric hyperbolic system},
\newblock Proc. Roy. Soc. Lond. A {\bf 378}, 401 (1981).

\bibitem{Fri81a}
H.~Friedrich,
\newblock {\em On the regular and the asymptotic characteristic initial value
  problem for {Einstein}'s vacuum field equations},
\newblock Proc. Roy. Soc. Lond. A {\bf 375}, 169 (1981).

\bibitem{Fri86a}
H.~Friedrich,
\newblock {\em On purely radiative space-times},
\newblock Comm. Math. Phys. {\bf 103}, 35 (1986).

\bibitem{Fri88}
H.~Friedrich,
\newblock {\em On static and radiative space-times},
\newblock Comm. Math. Phys. {\bf 119}, 51 (1988).

\bibitem{Fri98a}
H.~Friedrich,
\newblock {\em Gravitational fields near space-like and null infinity},
\newblock J. Geom. Phys. {\bf 24}, 83 (1998).

\bibitem{FriKan00}
H.~Friedrich \& J.~K\'{a}nn\'{a}r,
\newblock {\em Bondi-type systems near space-like infinity and the calculation
  of the {N}{P}-constants},
\newblock J. Math. Phys. {\bf 41}, 2195 (2000).

\bibitem{Joh91}
F.~John,
\newblock {\em Partial differential equations},
\newblock Springer, 1991.

\bibitem{Kan96b}
J.~K\'{a}nn\'{a}r,
\newblock {\em On the existence of $\mbox{{C}}^\infty$ solutions to the
  asymptotic characteristic initial value problem in general relativity},
\newblock Proc. Roy. Soc. Lond. A {\bf 452}, 945 (1996).

\bibitem{NewPen62}
E.~T. Newman \& R.~Penrose,
\newblock {\em An approach to gravitational radiation by a method of spin
  coefficients},
\newblock J. Math. Phys. {\bf 3}, 566 (1962).

\bibitem{PenRin84}
R.~Penrose \& W.~Rindler,
\newblock {\em Spinors and space-time. {V}olume 1. {T}wo-spinor calculus and
  relativistic fields},
\newblock Cambridge University Press, 1984.

\bibitem{SacBer58}
R.~K. Sachs \& P.~G. Bergmann,
\newblock {\em Structure of particles in linearized gravitational theory},
\newblock Phys. Rev. {\bf 112} (1958).

\bibitem{Sze78}
G.~Szeg\"{o},
\newblock {\em Orthogonal polynomials}, volume~23 of {\em {A}{M}{S} {C}olloq.
  {P}ub.},
\newblock A{M}{S}, 1978.

\bibitem{Tay96}
M.~E. Taylor,
\newblock {\em Partial differential equations {I}},
\newblock Springer, 1996.

\bibitem{Val98}
J.~A. Valiente~Kroon,
\newblock {\em Conserved Quantities for polyhomogeneous spacetimes},
\newblock Class. Quantum Grav. {\bf 15}, 2479 (1998).

\bibitem{Val99b}
J.~A. Valiente~Kroon,
\newblock {\em A Comment on the Outgoing Radiation Condition and the Peeling
  Theorem},
\newblock Gen. Rel. Grav. {\bf 31}, 1219 (1999).

\bibitem{Val99a}
J.~A. Valiente~Kroon,
\newblock {\em Logarithmic {Newman}-{Penrose} Constants for arbitrary
  polyhomogeneous spacetimes},
\newblock Class. Quantum Grav. {\bf 16}, 1653 (1999).

\bibitem{Val00}
J.~A. Valiente~Kroon,
\newblock {\em On the existence and convergence of polyhomogeneous expansions
  of zero-rest-mass fields},
\newblock Class. Quantum Grav. {\bf 17}, 4365 (2000).

\bibitem{Val00a}
J.~A. Valiente~Kroon,
\newblock {\em Polyhomogeneity and zero-rest-mass fields with applications to
  {Newman}-{Penrose} constants},
\newblock Class. Quantum Grav. {\bf 17}, 605 (2000).

\bibitem{Wal84}
R.~M. Wald,
\newblock {\em General Relativity},
\newblock The University of Chicago Press, 1984.

\end{thebibliography}

%

\end{document}